\documentclass[pra,twocolumn,groupedaddress]{revtex4-2}
\usepackage[english]{babel}
\usepackage[T2A]{fontenc}
\usepackage{lmodern}
\usepackage{graphicx}
\usepackage{amsmath}
\usepackage{amsfonts}
\usepackage{amssymb}
\usepackage{amsthm}
\usepackage{mathtools}
\usepackage{dsfont}
\usepackage{microtype}
\usepackage{tempora}
\usepackage[usenames]{color}
\usepackage{colortbl}
\usepackage{xcolor}
\usepackage{ulem}
\usepackage{easyReview}
\usepackage{array}
\usepackage{multirow}

\usepackage{hyperref}
\hypersetup{colorlinks=true, linkcolor=blue, citecolor=blue, urlcolor=blue}

\begin{document}
	
	\def\BE{\begin{equation}}
		\def\EE{\end{equation}}
	\def\BA{\begin{array}}
		\def\EA{\end{array}}
	\def\BEA{\begin{eqnarray}}
		\def\EEA{\end{eqnarray}}
	\def\nn{\nonumber}
	\def\ra{\rangle}
	\def\la{\langle}
	\def\p{{\bf p}}
	\def\q{{\bf q}}
	\def\A{{\bf A}}
	\def\x{{\bf x}}
	\def\S{{\bf S}}
	\def\O{{\bf \Omega}}
	\def\I{{\bf I}}
	\def\U{{\bf U}}
	\def\V{{\bf V}}
	\def\Z{{\bf Z}}
	\def\d{\partial}
	
	\title{Optical Schrödinger cat states generation using cubic phase resource state and beamsplitter}
	\author{A.~V.~Baeva}
	\affiliation{Saint Petersburg State University, 7/9 Universitetskaya nab., Saint Petersburg, 199034 Russia}
	\author{A.~S.~Losev}
	\affiliation{Saint Petersburg State University, 7/9 Universitetskaya nab., Saint Petersburg, 199034 Russia}
    \affiliation{Saint Petersburg State Marine Technical University, 3 Lotsmanskaya str., Saint Petersburg, 190121 Russia}
	\author{I.~V.~Sokolov}
	\affiliation{Saint Petersburg State University, 7/9 Universitetskaya nab., Saint Petersburg, 199034 Russia}
	
	\begin{abstract}
        Squeezed Schrödinger cat states are a valuable resource for quantum error correction and quantum computing. In this paper, we investigate the gate for generating such states in the optical regime. Our scheme is based on the entanglement between an arbitrary (in general) signal and a resource non-Gaussian light fields on an asymmetric beamsplitter, followed by homodyne measurement. 
        The resource field is preconditioned in the cubic phase state. In contrast to the previously considered gates that use the same resource state and QND entangling operation, the beamsplitter-based scheme offers a straightforward possibility for obtaining squeezed Schrödinger cat states with the desired degree of squeezing. 
        Along with the exact description of the gate, we perform here the semiclassical analysis of the gate operation we have introduced previously. This allows us to demonstrate clearly the principle of gate operation and to reveal in a simple and visual form the significant statistical properties of the output state, such as the squeezing ratio and the specific deformations of the output cat state components.
        Comparative analysis of the gate efficiency is presented for different resource states, that is, the cubic phase state and the Fock state. The gate parameters that ensure the generation of squeezed Schrödinger cat states with a needed fidelity and probability are evaluated.

        \end{abstract}
	
	\date{\today}
	
	\maketitle
	
	\section{Introduction}
    
    The generation of Schrödinger cat-like optical states which are superpositions of macroscopically distinguishable states of light is of considerable interest \cite{Yurke1986, Gerry-Knight-1997, Cochrane1999}. These states have applications in fundamental problems, such as the problem of degree of macroscopicity in quantum systems \cite{Leggett2002, Arndt2014}, and in applied ones, including quantum metrology \cite{Gilchrist2004, Giovannetti2011} and quantum computing \cite{Gilchrist2004, Ralph2003, Lund2008}. 
    The Schrödinger cat states have proven to be effective in correcting common types of errors in quantum error correction protocols \cite{Lund2008, Guillaud2023}. 
    Error correction codes based on the use of squeezed cat states, represented as a linear superposition of mutually displaced squeezed states, have the potential to improve robustness against dephasing and to partially correct errors caused by photon loss \cite{LeJeannic2018, Schlegel2022}.

    There are a number of approaches for implementing optical Schrödinger cat states. One possibility is based on the evolution of the initial state subject to a strong nonlinear interaction. This can result in distinct evolution of the canonical variables of the input field in different regions of the phase plane \cite{Yurke1986, RiveraDean2021}. 
    Another approach involves the quantum entanglement in the input state and the subsequent measurement in one or more of the output channels. Knowledge about the target oscillator variables that is compatible with the measurement outcome may contain multivalued  information about the output state variables in the signal channel of the system. This leads to two (or more) ``copies'' of the target state on the phase plane of the output state. 
    
    This feature can arise due to the non-Gaussianity in the system prior to Gaussian quadrature amplitude measurement \cite{Ourjoumtsev2007}, where the Fock resource state is used and the ``two-headed'' cat state is produced. The non-Gaussian nature of the measured quantity manifests itself in protocols involving the subtraction of photons from Gaussian resource state  \cite{Yurke1990, Korolev2024, Ourjoumtsev2006, Takase2022}, and for photon-number resolving measurements utilizing parity detectors \cite{Thekkadath2020}.  
   Using iterative methods and more complex measurement procedures, one can create Schrödinger cat states and also increase their amplitude \cite{Lund2004, Sychev2017, Eaton2022, Winnel2024}. The non-Gaussian small-amplitude cat states can simultaneously serve as both signal and resource input fields \cite{Lund2004, Sychev2017}, thus ensuring the multivalued nature of the information about the output state variables. In a multichannel circuit \cite{Winnel2024}, the incremental use  of non-Gaussian Fock resource state and non-Gaussian measurement in the form of projection onto squeezed Fock states allows to create various superpositions at the circuit output, including ``four-headed'' Schrödinger cat states.  Note that one can easily address properties of such states using semiclassical approach.

    As demonstrated previously \cite{Sokolov2020, Masalaeva2022, Baeva2023, Baeva2024}, the non-Gaussian cubic phase state can be used as a resource not only in gates designed for nonlinear control of  phase of the output signal, but also for the generation of Schrödinger cat states. That is, the lowest-order non-Gaussian resource: a cubic phase state is sufficient for this goal. In these works, we have developed an efficient and illustrative method that allows one to describe the emerging Schrödinger cat states. This method is based on the semiclassical interpretation of Heisenberg representation, when the measured observables  are treated as  c-numbers. The semiclassical mapping of signal quadrature amplitudes can arise as multivalued, allowing one to determine the number of cat components on the phase plane. In some cases \cite{Sokolov2020, Masalaeva2022, Baeva2023, Baeva2024}, one can evaluate their position, shape, and possible deformation with good fidelity. 
        
    As shown in \cite{Masalaeva2022}, a gate based on the entangling $C_{Z}$ operation and the resource cubic phase state can prepare Schrödinger cat states from arbitrary input states, for example, from Fock states with a few photons. We have found \cite{Baeva2023} the conditions under which the use of the real cubic phase state (i.e. obtained under the condition of limited initial squeezing)  allows for one to obtain the needed output states with the required fidelity. In general, a variety of quantum states that allow a multivalued semiclassical representation of observables can be used as a resource, e. g., the photon-number states. A comparison \cite{Baeva2024} of gates based on two different resource states (the cubic phase state and the Fock state) allowed us to identify the conditions under which these gates provide the necessary output. 

    In this paper, we investigate a gate that uses the cubic phase state with a realistic initial squeezing as a resource, as mentioned previously, but entangles the input light fields on a beamsplitter with arbitrary reflection and transmission coefficients. It is known \cite{Ourjoumtsev2007, Winnel2024} that in the gate using Fock resource state the squeezed Schrödinger cat states occur at the output of the beamsplitter. We show here that controllable squeezed states of this type also emerge with the cubic phase state used as a resource. We perform the  semiclassical analysis that predicts the squeezing and deformation of the output state components for both resource states in a simple and illustrative way. These results are in a good agreement with the exact Wigner functions of the output states and with the corresponding measure (fidelity). This allows one to choose the proper gate parameters that are compatible with the output squeezed Schrödinger cat states with needed properties with high quality and probability.      		
            
	
    \section{Gate for squeezed Schrödinger cat state generation} 
    \label{GateDef}

    In the scheme whose operation is illustrated in Fig.~\ref{figBS}, two quantum light fields are mixed at an asymmetrical beamsplitter. The states of the oscillators at the input are called the signal and the resource (ancilla), respectively. The scheme is intended to prepare a superposition of squeezed ``copies'' of the signal state at the target output.      
    In general, the signal field can be prepared in an arbitrary state that occupies a restricted area on the phase plane. The cubic phase state of the resource ensures two-valued information about the field quadrature amplitudes at the gate output after the beamsplitter and homodyne measurement of the ancilla. 
   
    Then, as shown in \cite{Sokolov2020, Masalaeva2022, Baeva2023, Baeva2024}, at the gate output a quantum state emerges that is represented by the superposition of two ``copies'' of the signal in the phase plane. 
    In configurations similar to our scheme, it is necessary that the resource state be non-Gaussian. Such states include, for instance, Fock states with a well-defined number of photons, the generalised squeezed states produced by higher-order nonlinear interactions \cite{Braunstein1990, Zelaya2018}, the cat or grid states, etc. The non-Gaussian cubic phase state was recently obtained in the microwave range \cite{Kudra2022, Eriksson2024}.

    To obtain the output state as a superposition of two mutually displaced squeezed vacuum states, we assume the signal state be vacuum with the following wave function,	
			\BE
			\label{vac_state}
			\psi_{1}(x_{1}) = \frac{1}{\pi^{1/4}}e^{-x_{1}^{2}/2}.
			\EE
    Here, the signal and the ancilla are labeled by index 1 and 2, respectively.
    
    We define the cubic phase state as the vacuum state initially squeezed  along the momentum axis, with the wave function
                \BE
                \label{SqDef}
			\psi^{(sq)}_{2}(x_{2}) = \frac{s^{1/2}}{\pi^{1/4}}e^{-\frac{1}{2}s^{2}x_{2}^{2}}.
                \EE
    The squeezing parameter $s$ is chosen such that $s<1$, and $x_{2}$ is the coordinate of the second oscillator. Then, the state evolves with the cubic Hamiltonian $\hat{H}=~-\gamma_{0} \hat{q}_{2}^{3}$ with the nonlinearity parameter $\gamma_{0}$. The state vector of the resource oscillator becomes
                \BE\label{cubstate}
			|\psi_{2}^{(cub)}\ra = e^{i\gamma\hat{q}^{3}}|\psi^{(sq)}_{2}\ra\nn, 
			\EE
    where the wave function of the squeezed cubic phase state is
			\BE\label{cubphase}
			\psi^{(cub)}_{2}(x_{2}) = \psi^{(sq)}_{2}(x_{2})e^{i\gamma x_{2}^{3}}.
			\EE
    Here $\gamma = \gamma_{0}t$, and $t$ is the evolution interval. The state vector of the system at the beamsplitter input is as follows,
			\BE\label{InputState}
			|\psi_{12}^{(in)}\ra = \iint  dx_{1}dx_{2}\;\psi_{1}(x_{1})\psi^{(cub)}_{2}(x_{2})|x_{1}x_{2}\ra.
			\EE
        The beamsplitter scattering matrix $S$ is taken in the form
				\BE
			\label{Smatr}
			S=
			\begin{pmatrix}
			\rho & -\tau \\
			\tau & \rho
			\end{pmatrix},
			\EE
        with $S$ unitary. The real reflection and transmission coefficients denoted by $\rho$ and $\tau$, respectively, satisfy $\rho^{2} + \tau^{2} = 1$, with no phase shifts in the channels.

        The output wave function of the beamsplitter is related to the input one as
				\BE
			\label{STransform}
			\psi^{(BS)}_{12}(\vec{x}) = \psi_{12}^{(in)}(S^{-1}\vec{x}),
			\EE
        where $\vec{x}=(x_{1},x_{2})^{T}$ is the vector of the wave function arguments. In the Heisenberg picture, for a beamsplitter with a real scattering matrix the canonical variables transformation is performed with the same matrix $S$, which in the latter case is block of the symplectic matrix.
        
        Thus,
	       \BE	
           \label{PsiOut}
		     \psi_{12}^{(BS)}(x_{1},x_{2})=\psi_{1}(\rho x_{1}+\tau x_{2})\psi_{2}^{(cub)}(-\tau x_{1}+\rho x_{2}).
		     \EE

        Next, a homodyne measurement of momentum quadrature of the field in the second channel is performed, and the measurement outcome $y_{m}$ with some probability is obtained. This corresponds to the observation of the momentum operator $\hat{p}_{2}^{(out)}$ eigenvector,
                \BE\label{Eigenvector}
			|y_{m}\ra = \frac{1}{\sqrt{2\pi}} \int dx_{2} e^{i x_{2} y_{m}}|x_{2}\ra.
		    \EE
        As a result of the measurement, the quantum state (\ref{PsiOut}) is projected onto the eigenstate \eqref{Eigenvector} of the homodyne detector,
            \BE
			|\widetilde{\psi}^{(out)}\ra = \la y_{m}|\psi^{(BS)}_{12}\ra. \nn
		\EE
        The unnormalised wave function of the reduced state is given by
            %
            %
            \BEA\label{CatWavefunction}
			\begin{split}
			\widetilde{\psi}^{(out)}_{y_{m}}(x_{1})= \frac{1}{\sqrt{2\pi}}  \int\limits_{-\infty}^{+\infty} dx_{2} \; \psi_{1}(\rho  x_{1} + \tau  x_{2}) \times \\ 
			\times \psi^{(cub)}_{2}(-\tau  x_{1} + \rho  x_{2})  e^{-i x_{2} y_{m}}.
			\end{split}
			\EEA
        Here, the norm of the reduced state gives the probability density to obtain a Schrödinger cat state at the gate output,
                \BE\label{probability}
			P(y_{m})= \la \widetilde{\psi}^{(out)}_{y_{m}}|\widetilde{\psi}^{(out)}_{y_{m}}\ra = \int dx \; |\widetilde{\psi}^{(out)}_{y_{m}}(x)|^{2},
			\EE
        which depends on the parameters of the scheme and on the measurement outcome in the second channel.

        Finally, the explicit form of the normalized output wave function is found to be
                \BE\label{PsiInt}
			\psi^{(out)}_{y_{m}}(x) = \sqrt{\frac{1}{P(y_{m})} }\sqrt{\frac{s}{2 \pi^{2}} } e^{d(x)} \! \int\limits_{-\infty}^{+\infty}\!\! dx_2 \, e^{i\left[a x_{2}^{3}+b(x) x_{2}^{2}+c(x) x_{2}\right]},
			\EE
        where we define 
				\BEA
			\begin{split}
			&a = \rho ^{3} \gamma,  \\
			&b(x) = i(\rho ^{2} s^{2} + \tau ^{2})/2 - 3 \rho ^{2} \tau  \gamma x ,  \\
			&c(x) =  i \rho  \tau  x - i \rho  \tau  s^{2} x - y_{m} + 3  \rho  \tau ^{2} \gamma x^{2},  \\
			& d(x)=- i\tau ^{3}\gamma x^{3}-(\rho ^{2}+\tau ^{2} s^{2})x^{2}/2, 
			\end{split}
			\EEA
        and $x$ is a shorthand for $x_{1}$.
                
         The integral \eqref{PsiInt} is evaluated analytically using the useful expression from \cite{vallee2010airy},
			\BEA
			\begin{split}
			&\int\limits_{-\infty}^{+\infty} \! dx_{2}\; \exp \left[ i (a x_{2}^{3} +  b(x) x_{2}^{2} + c(x) x_{2}) \right] = \frac{2\pi}{(3a)^{1/3}}  \times \\ 
			&\exp\! \left[ - i \frac{b(x)}{3 a} \left( c(x) - \frac{2b^{2}(x)}{9 a} \right) \right] \text{Ai}\! \left[ \frac{ c(x) - b^2(x)/3 a }{(3a)^{1/3}} \right].
			\end{split}
			\EEA
        Here \text{Ai}(x) is the Airy function of the first kind.
	
        With the assumption $\rho ^{2}+\tau ^{2} = 1$ of the beamsplitter unitarity, the gate output wavefunction (\ref{PsiInt}) finally becomes
				\BEA \label{OutputWF}
			\begin{split}
			&	\psi^{(out)}_{y_{m}}(x) = \frac{\sqrt{2s/P(y_{m})}}{\rho (3\gamma)^{1/3} } \exp\left[-\frac{x^{2}}{2\rho ^{2}}+ 2 \gamma \Phi^{3}-\frac{\Phi y_{m}}{\rho }\right] \times \\
			&	 \exp\left[\frac{i\tau }{\rho }\left(\frac{\Phi}{\rho }-y_{m}\right)x\right] \text{Ai}\left[ \frac{  i \tau x / \rho + 3\rho \gamma \Phi^{2}-y_{m} }{\rho (3\gamma)^{1/3}} \right],
			\end{split}
			\EEA
	where $\Phi = \left( \tau ^{2}/\rho ^{2}+ s^{2}\right)/6\gamma$.

	
    \section{Evolution of quadrature amplitudes in the semiclassical approximation}
    \label{sec_semicl}
        Consider evolution of quadrature amplitudes of the input fields starting from the Heisenberg picture. This will allow us to demonstrate explicitly how the superposition of two (or more in a general case) copies of the input signal appears in the output state, and why do we refer to the output state as the Schrödinger cat state.
        
        An essential feature of our approach is that the operator of a measured system variable is replaced with the c-number measurement outcome. This can make commutation relations for canonical variables in the Heisenberg picture inaccurate and render the theoretical picture to the semiclassical approximation.

        Generally, for the multivariable non-Gaussian states, an implied consequence of a measurement may be a multivalued information about the variables of the target oscillator.  It was demonstrated    \cite{Sokolov2020, Masalaeva2022, Baeva2023, Baeva2024}, that the semiclassical approach is able to effectively predict statistical properties of generated quantum states, including Schrödinger cat states. In what follows, we show that this method works well also for the gate using resource cubic phase state, beamsplitter and homodyne measurement. 
        
        The form of the output state, the position of its components in the phase plane, and possible deformations are effectively evaluated. This is of use in determining the optimal gate parameters needed to generate squeezed Schrödinger cat states with the desired properties.

\begin{figure}[t]
	\begin{center}
		\includegraphics[width=0.95\columnwidth]{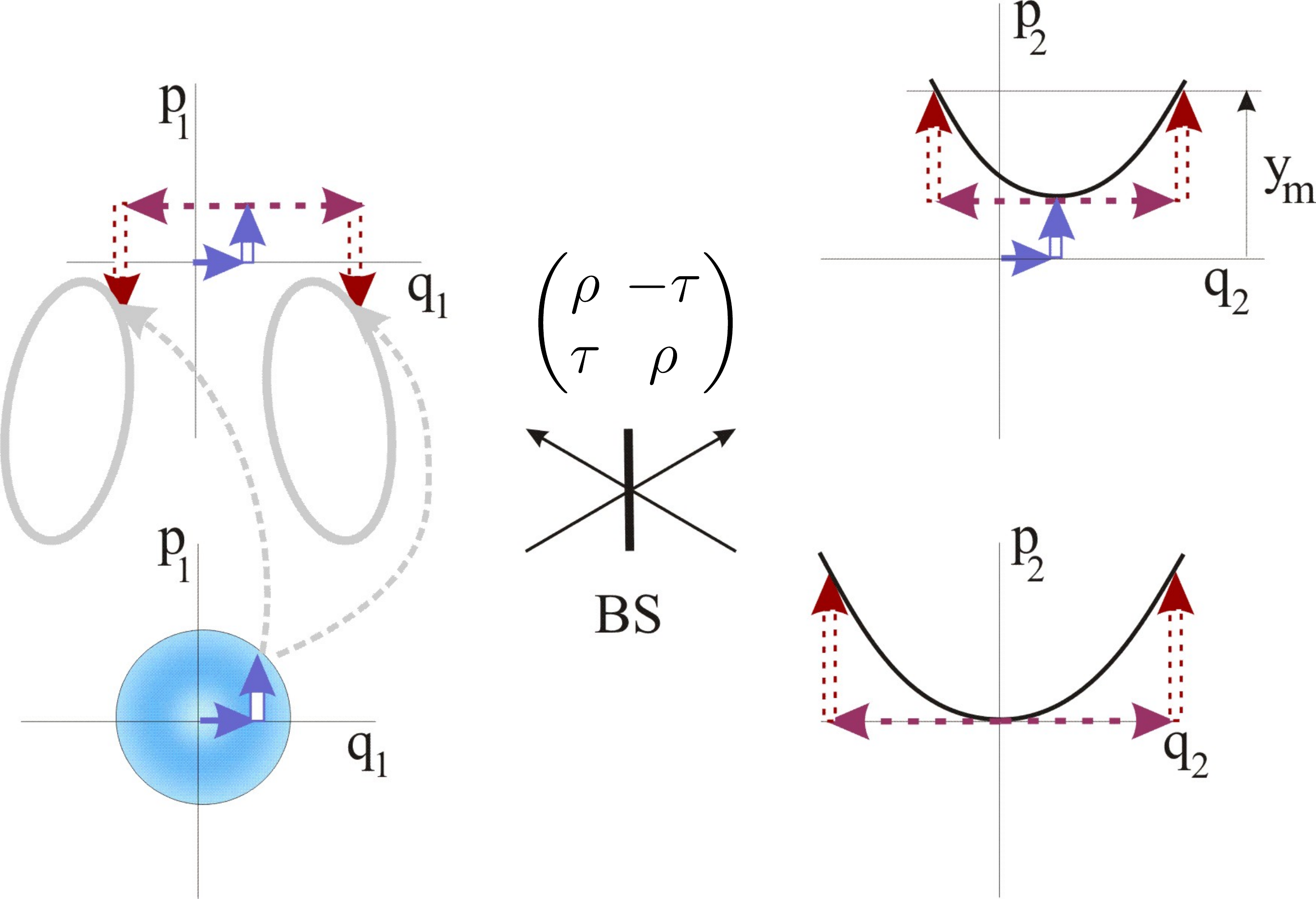}	
        \end{center}
	\caption{ Semiclassical mapping of quadrature amplitudes of the signal and the resource states as a result of the gate operation for perfect initial squeezing of the resource. The randomly chosen initial amplitudes of the input fields are mixed on a beamsplitter with the scattering matrix (\ref{Smatr}). After mixing, the momentum of the resource oscillator is measured with the outcome $y_{m}$, and a reduced quantum state in the signal channel emerges. In our pictorial representation, two intersections of the measurement line with the state curve specify two sets of quadrature amplitudes compatible both with the state preparation and with the measurement outcome. Due to the entanglement, the information about these two sets of amplitudes is imprinted in the output state in the first channel, and a Schrödinger cat-like state is created.
    }
	\label{figBS}
    \end{figure}

        The in - out mapping of an arbitrarily chosen point in the phase plane of the signal field is shown in Fig. \ref{figBS}.
    The initial state of the signal field is vacuum (note that other states can be treated \cite{Masalaeva2022} using a similar approach). Its Wigner function is represented by a circle in the phase plane. The resource field \eqref{cubphase} is prepared from the perfect momentum-squeezed state which has undergone evolution in presence of the cubic Hamiltonian $\hat{H}~=~-\gamma_{0}\hat{q}_{2}^{3}$. The Heisenberg equation for the momentum of the resource is
            \BE
            \dot{\hat{p}}_{2}	=i[\hat{H},\hat{p}_{2}]= -i\gamma_{0} [\hat{q}^{3}_{2} ,\hat{p}_{2}] = 3 \gamma_{0} \hat{q}_{2}^{2}, 
	    \EE
    etc., where  $[q_{2},p_{2}]= i\hbar $, $\hbar=1$, and $\hat{q}_{2}, \hat{p}_{2}$ are the coordinate and momentum operators of the resource field. We arrive at 
		\BEA
            \label{NonG_amplitudes}
	    \begin{split}
		&\hat{q}_{1}= \hat{q}_{1} (0),\\
		&\hat{p}_{1} = \hat{p}_{1} (0),\\ 
		& \hat{q}_{2}^{(cub)} = \hat{q}_{2} (0),\\
		&\hat{p}_{2}^{(cub)} = \hat{p}_{2} (0) + 3\gamma \hat{q}_{2}^{2} (0),
	    \end{split}
	    \EEA
        Here, $\hat{q}_{j} (0), \hat{p}_{j} (0)$, with $j=1,2$, are quadrature amplitudes of the input fields before cubic phase transformation  and $\gamma = \gamma_{0} t$.

        Next, the light fields are entangled at the beamsplitter with the real scattering matrix $S$, which ensures that coordinates and momenta are not mutually mixed. The in - out relation of the scheme at this stage reads
	   \BEA \label{PQEnt}
	   \begin{split}
	   &\hat{q}_{1}^{(ent)} = \rho \hat{q}_{1} (0) - \tau  \hat{q}_{2}(0),\\
	   &\hat{p}_{1}^{(ent)} = \rho  \hat{p}_{1} (0) - \tau \left[\hat{p}_{2}(0)+3\gamma(\hat{q}_{2}(0))^{2}\right],\\
	   &\hat{q}_{2}^{(ent)} = \tau \hat{q}_{1} (0) +\rho  \hat{q}_{2}(0),\\
	   &\hat{p}_{2}^{(ent)} = \tau  \hat{p}_{1} (0) +\rho \left[\hat{p}_{2}(0)+3\gamma(\hat{q}_{2}(0))^{2}\right].
	   \end{split}
	   \EEA
       Finally, a homodyne measurement of momentum is performed in the second output channel of the beamsplitter, with a random measurement outcome $p_{2}^{(m)} = y_{m}$. By substituting the c-number $y_m$ for the momentum quadrature operator, $p_{2}^{(ent)} \rightarrow p_{2}^{(m)} = y_{m}$, we invoke a semiclassical approximation. Then, in (\ref{PQEnt}) we express  the coordinate $q_{2}(0)$ in terms of other observables. Two possible values of the input coordinate of the resource field that are specified by the measurement, $\hat q_{2}(0) \to q_2^{(m)}$, arise, 
            \BEA	 \label{PQMeas}
		q_{2}^{(m)} = \tau q_{1} (0) \pm\sqrt{\rho /3\gamma } \left[y_{m} - \tau  {p}_{1}(0)-\rho {p}_{2}(0)\right]^{1/2},
		\EEA
         as indicated in Fig. \ref{figBS} by horizontal red dashed arrows. Eventually, the two-valued semiclassical in - out relation of the gate (that is, for its signal field) is found in the form
            \BEA \label{shifts}
	   \begin{split}
		&{q}_{1}^{(out)}=\rho {q}_{1}(0) \mp \frac{\tau }{\sqrt{3\gamma \rho }} \left[y_{m} - \tau  {p}_{1}(0)-\rho {p}_{2}(0)\right]^{1/2},\\
		&{p}_{1}^{(out)}= \frac{1}{\rho }\left[{p}_{1}(0) - \tau  y_{m}\right].
	   \end{split}
	   \EEA
         In the last expression for the output state momentum, the squeezing parameter $1/\rho$ appears, which manifests squeezing of both output state components performed by the gate. 

        Since our analysis is performed in the limit of infinite initial squeezing in momentum of the resource state, as illustrated in Fig.\ref{figBS}, we assume $p_{2}(0) \to 0$. Consequently, the expression \eqref{shifts} predicts in the semiclassical approximation a two-valued mapping of a randomly chosen initial point in the phase plane of the signal field to a double set of separated final points.
        

        The semiclassical description of the evolution of input quadrature amplitudes provides a clear visual representation of the emergence of two squeezed copies of the target state at the gate output.  It also predicts the position of every copy in the phase plane, as well as the distance between copies depending on the gate parameters and the measurement outcome.  Furthermore, it allows us to evaluate the deformation of the output state components. 
        
        The half-spacing in coordinate between two destinations of an arbitrary initial point is
            \BE \label{DeltaQ}
	\Delta q = \frac{\tau }{\sqrt{3\gamma \rho }} \left[y_{m} - \tau  {p}_{1}(0)
	\right]^{1/2},
		\EE
        as follows from \eqref{shifts}. Since the coordinate of a destination also depends on the momentum of the initial point, a nonlinear deformation arises. For $\tau p_1(0) \ll y_m$, one can linearize $q_1^{(out)}$ in the momentum ${p}_{1}(0)$, and an opposite linear shearing deformation of the copies arises. 
        
        As seen from Fig. \ref{figBS}, these deformations are explicitly due to the shape of the resource state, the cubic phase state in our model.  This allows one to approximate the Schr\"odinger cat state state at the output of the gate as a superposition of two mutually displaced, squeezed and skewed copies of the target state. Note that the semiclassical description of the gate operation makes it possible to infer the shape of the output state in the phase plane, but does not provide information about the relative phase between the copies.


    \section{Perfect squeezed cat state and gate output: fidelity and probability} \label{secIV}
    
    \subsection{Perfect cat state and fidelity}
    
    To be specific, we define a perfect squeezed Schrödinger cat state that can be used in quantum error correction protocols (e.~g. \cite{Schlegel2022, Korolev2024}) as a superposition of two symmetrically shifted in coordinate and squeezed vacuum states. These states are located at $\alpha = \pm\alpha’ + i\alpha''$ in the phase plane, and their squeezing is defined by the parameter $s' = 1 /\rho$, as follows from our semiclassical analysis (see (\ref{shifts})). 
    Here $\alpha'= \tau \sqrt{y_{m}}/\sqrt{3\gamma \rho}$ and $\alpha'' = -\tau y_{m}/\rho$. We do not include into our definition of a perfect cat state the linear and nonlinear deformation of the generated copies of the signal state, seen from (\ref{shifts}). 
    
    The wavefunction of a perfect cat state is
            \BEA \label{IdCat}
	    \begin{split}
	    \psi^{(id)}_{cat}(x) = N_{c}\left( \exp \left[i\frac{\theta}{2}-s'^{2}\frac{(x-\alpha')^{2}}{2}+i\alpha'' x\right]+\right.\\
	    \left.	+\exp \left[-i\frac{\theta}{2}-s'^{2}\frac{(x+\alpha')^{2}}{2}+i\alpha'' x \right]\right),
	    \end{split}
	    \EEA
        where $\theta$ is the relative phase between the components of cat state, and $N_{c} =  \sqrt{2(1+\cos(2\theta)e^{s'^{2}\alpha'^{2}})}^{-1}$ is the normalization factor. 
        It is challenging to explicitly extract the relative phase $\theta$ from the exact wavefunction (\ref{OutputWF}) of the output state. One approach to this problem is to calculate the fidelity between the output state and the perfect cat state,
            \BE
            \label{FidPerf}
		F = \left|\int dx\; \psi_{cat}^{(id)*}(x)\psi^{(out)}(x)\right|^{2},
		\EE
        for any relative phase. Since a possible standard set of cat states used as a basis for quantum information protocols consists of even and odd states, in what follows we consider an odd cat state, and set $\theta$ to $\pi$. An analysis of an even cat state may be performed along the same lines. 

        In this section, we analyze the fidelity (\ref{FidPerf}) in dependence of the measurement outcome $y_m$, the cubic nonlinearity parameter $\gamma$, and the initial resource squeezing $s$, see (\ref{SqDef}).
        To be specific, we consider here an asymmetric beamsplitter with $\rho = 1/2, \tau =\sqrt{3}/2$, which corresponds to squeezing $s' = 2$ of the cat state components. The value of $\alpha'$ is set to $\alpha'=\Delta q  = 2.87$. This is done to facilitate a comparison of the efficiency of our cubic phase state-based gate with the gate that uses a Fock resource state at the input, with the number of photons $n=5$ (see Sec.~VI).	
        To make our analysis more realistic, we assume that the fidelity (\ref{FidPerf}) is evaluated between the output state of a gate with a real achievable in the microwave region  initial squeezing $s=0.2 \; (\approx 14 \;\text{dB})$, and a perfect cat state (\ref{IdCat}) inferred from the semiclassical analysis.
        
\begin{figure}[t]
		\begin{center}
			\includegraphics[width=0.95\columnwidth]{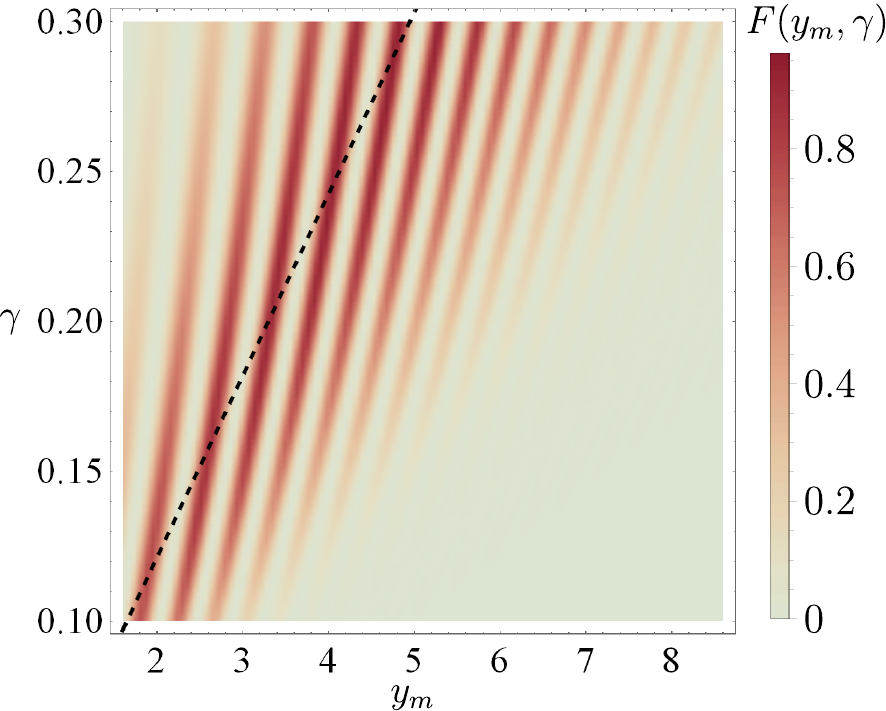}
		\end{center}
		\caption{Fidelity between the gate output state \eqref{OutputWF} and perfect odd cat state \eqref{IdCat} in dependence of the momentum measurement outcome $y_{m}$ and the cubic nonlinearity parameter $\gamma$. The reflection and transmission coefficients of an asymmetric beamsplitter are set to $\rho = 1/2$ and $\tau =\sqrt{3}/2$, respectively, and a realistic initial squeezing of the resource state of 14 dB $(s=0.2)$ is assumed. 
        The black dashed line specifies the set of parameters $\{y_{m},\gamma\}$ where the scheme generates cat states with a given half-distance \eqref{DeltaQ} between the centers of copies, which is fixed at  $\Delta q = 2.87$.} 
		\label{figPlotFidelity}
	\end{figure}

         Fig. (\ref{figPlotFidelity}) shows that there are ranges of parameters $\{y_m,\,\gamma\}$, where the state generated by the gate is close to the specified perfect odd cat state.

\begin{figure}[t]
		\begin{center}
			\includegraphics[width=0.8\columnwidth]{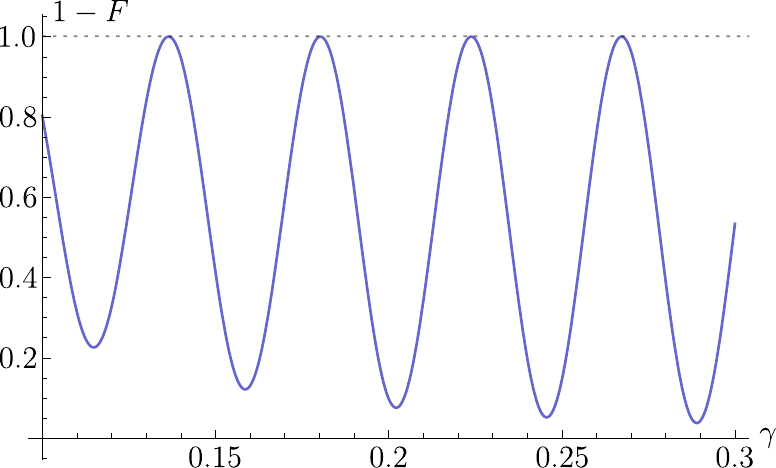}
		\end{center}
		\caption{Infidelity between the output state \eqref{OutputWF} with an initial squeezing of the resource state of 14 dB $(s=0.2)$ and the perfect odd Schrödinger cat state \eqref{IdCat}. Here 
        the beamsplitter parameters are $\rho = 1/2$, $\tau =\sqrt{3}/2$, and half-spacing between the copies is $\Delta q=2.87$.}
		\label{figInfidelity}
	\end{figure}

    The slice of the inverted fidelity distribution shown in Fig.~\ref{figPlotFidelity} (that is, of the infidelity $1 - F$) along the dashed line, where the half-spacing $\Delta q$ is fixed at $2.87$, is presented in Fig.~\ref{figInfidelity} as a function of the cubic nonlinearity parameter $\gamma$.
    After the value of $\gamma$ that ensures a necessary fidelity is specified using Fig.~\ref{figInfidelity}, one can choose an optimal value of the measurement output $y_{m} = \left(\sqrt{3\gamma\rho}/\tau\right)\Delta q$, as well as an acceptance interval $\Delta y_m$. 
	\begin{figure}[h]
		\begin{center}
			\includegraphics[width=0.8\columnwidth]{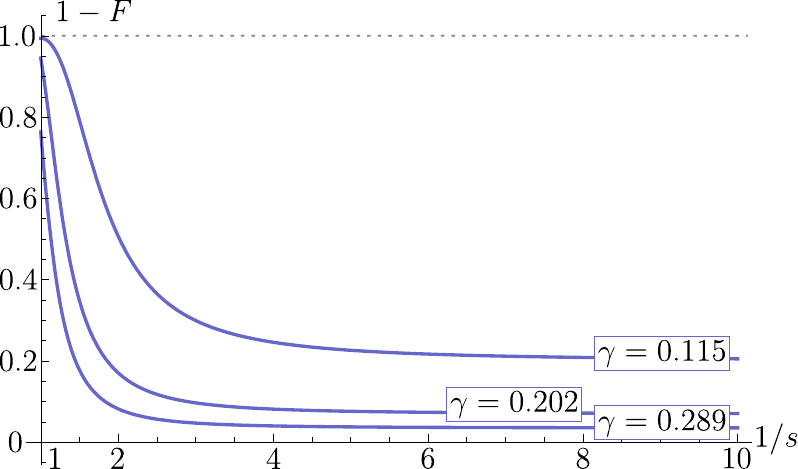}
		\end{center}
		\caption{Infidelity $1 - F$ plotted as a function of the initial squeezing $s$ of the resource cubic phase state for a set of optimal values of cubic nonlinearity, see Fig.~\ref{figInfidelity}. The beamsplitter parameters and the half-spacing between the copies are the same as in Fig.~\ref{figInfidelity}.}	\label{figInfidelitySqueezing}
	\end{figure}

        As follows from Fig.~\ref{figInfidelity}, the oscillating fidelity behavior indicates that the relative phase $\theta$ between the components of the output state depends on $\gamma$ and $y_{m}$ even when the half-spacing $\Delta q$ is fixed. The fidelity increases with the the nonlinearity parameter $\gamma$ of the resource state, as shown in Fig. \ref{figInfidelity}, \ref{figInfidelitySqueezing}. The latter can be easily clarified by referring the reader to the illustrative scheme in Fig. \ref{figBS}. As the nonlinearity parameter $\gamma$ increases, the slope of the parabola branches decreases. This slope is directly transferred to the shearing deformation of the ellipses, seen in Fig.~\ref{figBS} but not accounted for in the definition (\ref{IdCat}) of perfect output state. This effect is evidently associated with the ``shape'' of the resource state in the phase plane. 

        Regarding the effect of finite initial squeezing $s$ on gate fidelity, Fig.~\ref{figInfidelitySqueezing} shows that starting from some values of squeezing the main source of the output state imperfection is a shearing deformation rather than finite squeezing.

	
        \subsection{Probability of success}
        
       Consider the effect of the initial squeezing parameter $s$ on the probability density of the momentum measurement outcome $y_m$. Here we take for the cubic nonlinearity parameter $\gamma$ the values of 0.115, 0.202, and 0.289 that provide optimal fidelity within a certain range of $\gamma$, as follows from Fig.~\ref{figInfidelity}, and use the exact wavefunction (\ref{OutputWF}) of the output state. The beamsplitter parameters, the half-spacing between the copies, and squeezing of the copies are the same as above.
       
\begin{figure}[t]
		\begin{center}
			\includegraphics[width=0.8\columnwidth]{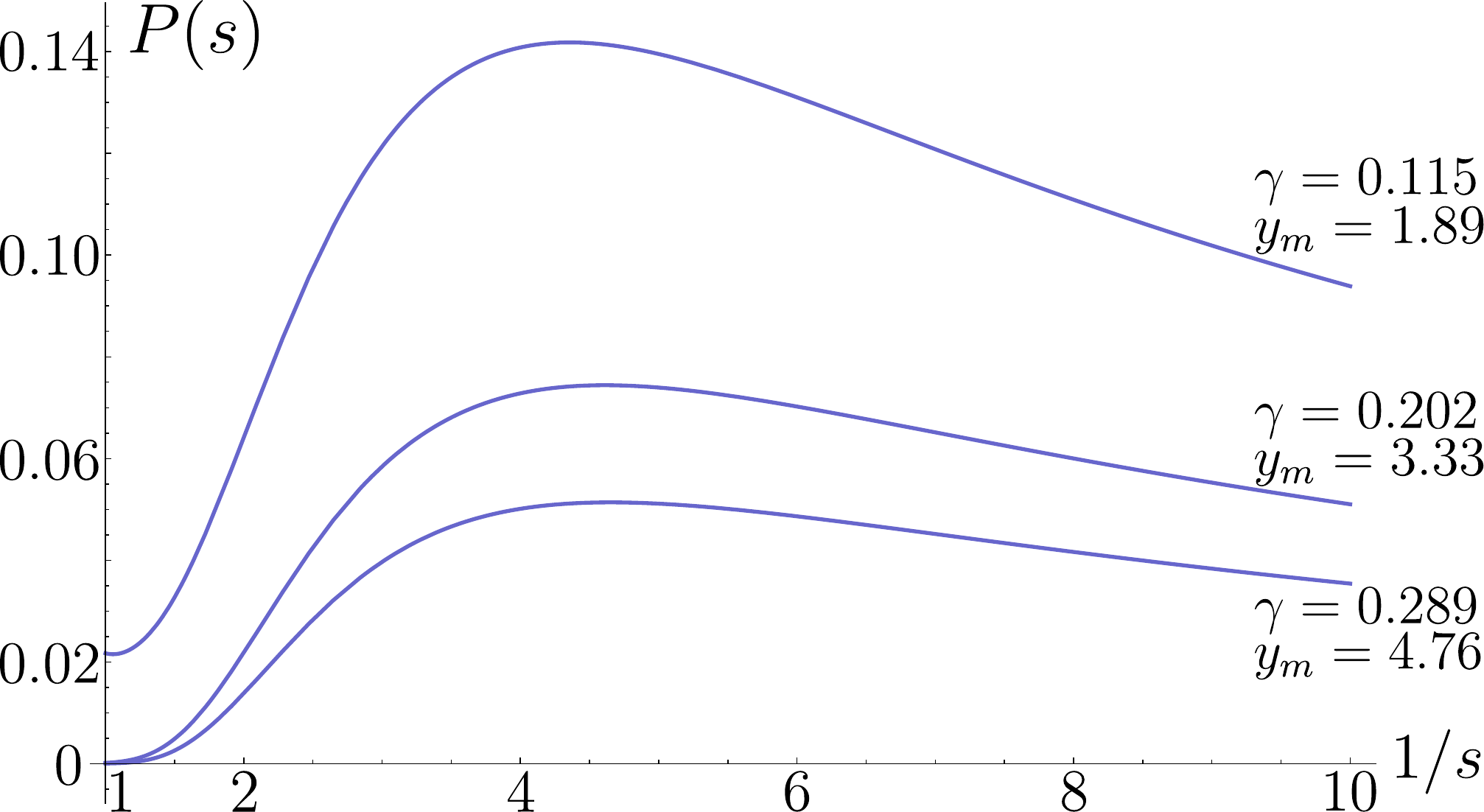}
		\end{center}
		\caption{Probability density of the momentum measurement outcome $y_m$ as a function of the initial squeezing of the resource cubic phase state for a set of optimal values of cubic nonlinearity and measurement outcome.}
		\label{figProbCatsFock}
	\end{figure}

        The probability density in dependence on the squeezing parameter $s$ of the resource state is plotted in Fig. \ref{figProbCatsFock}.
        In the absence of initial squeezing of the resource state, $s=1$, and for larger values of $\gamma$, the probability tends to zero since the measurement outcome lies outside the domain of existence of the resource state. For the chosen gate and cat state parameters, the maximum probability is achieved at initial squeezing of $\approx 12$ dB almost regardless of the resource state nonlinearity within a given range of values.
        
      As the initial squeezing increases, the resource state distribution in momentum covers a wider range. In a semiclassical picture, the distribution of cubic phase state in the phase plane approaches a perfect thin parabola, and the probability density vanishes. One can conclude that it is not necessary to achieve very large values of the initial resource state squeezing to maximize the probability of measurement success in our scheme.
	
	\section{Effect of beamsplitter parameters on output cat state squeezing and fidelity}

        In this section, we study the effect of an asymmetric beamsplitter parameters $\rho$ and $\tau$ (see the beamsplitter scattering matrix (\ref{Smatr})) on the output state components squeezing, and evaluate the achievable in every case fidelity. Note that the analysis of the preceding sections has been performed for $\rho =1/2; \; \tau = \sqrt{3}/2$, and hence for the squeezing of both copies of $s' = 2$. 
                
        As follows from our semiclassical analysis, the degree of squeezing of the output state components depends on the parameters of the beamsplitter (see \eqref{shifts}), $s' = 1/\rho$. Our next step is to construct a series of Wigner functions of the output states for a set of versions of our gate that differ from each other in the degree of output squeezing only. 
        These versions use the beamsplitter scattering matrix $S$ with $\{\rho=\sqrt{3}/2,\,\tau=1/2\}$, $\{\rho=1/\sqrt{2},\, \tau=1/\sqrt{2}\}$, $\{\rho=1/2,\,\tau=\sqrt{3}/2\}$, and $s'=2/\sqrt{3},\,\sqrt{2},\, 2$, correspondingly.  Here we fix the value of $\Delta q= 2.87$ of the half-spacing between  centers of the squeezed copies. This is done in order to make in Sec.~\ref{secComparison} a comparison with a similar scheme using Fock resource state with a fixed photon number of $n=5$ .
	\begin{figure*}[t]
		\centering
		\includegraphics[width=0.95\linewidth]{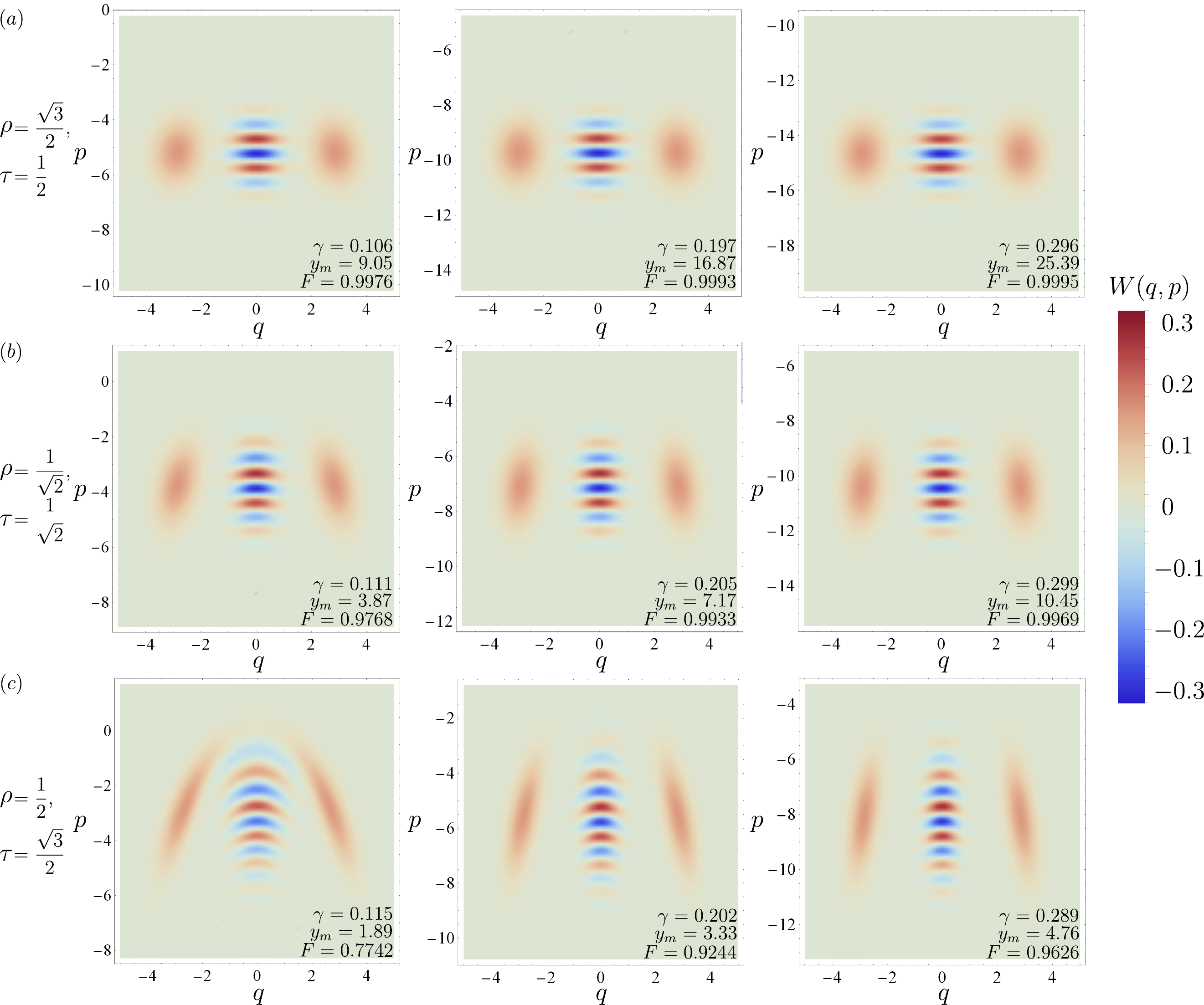}
		\caption{Wigner functions of the gate output states with fixed half-spacing between copies $\Delta q = 2.87$ and initial resource squeezing of 14 dB for different beamsplitter parameters and cubic nonlinearity $\gamma$. Here: $(a)$ - asymmetric beamsplitter with $\rho =\sqrt{3}/2,\,\tau =1/2$; $(b)$ - symmetric beamsplitter with $\rho = \tau = 1/2$; $(c)$ - asymmetric beamsplitter with $\rho =1/2,\, \tau =\sqrt{3}/2$.}
		\label{figCats}
	\end{figure*}

        Since the fidelity between the output state and the ideal odd Schrödinger cat state depends on the relative phase of the output state components, which in turn is related to the nonlinearity of the resource state, it is not possible to choose an arbitrary value of $\gamma$ for comparison. However, we can select values close to  $\gamma \approx 0.1;\; 0.2;\; 0.3$ that correspond to  local minima of infidelity within a specified range, (see Fig. \ref{figInfidelity}).	We have considered a wide range of values of the nonlinearity parameter $\gamma$ to demonstrate an increase in the fidelity of the output state as the nonlinearity of the resource state increases. However, by now only a value of $\gamma = 0.11$ has been experimentally achieved for the cubic phase state \cite{Eriksson2024}. As illustrated in Table \ref{Tab1}, each row corresponds to a gate using  beamsplitter with fixed reflection and transmission coefficients, and the resource state with an initial squeezing of 14 dB. For each gate, the optimal values of cubic nonlinearity parameters $\gamma$ close to $0.1;\; 0.2;\; 0.3$ are found, and the corresponding momentum measurement outcome $y_{m}$ is presented, as well as the fidelity between the output state \eqref{OutputWF} and perfect odd Schrödinger cat state \eqref{IdCat}.	
	 	
	\begin{table}[h]
		\caption{Parameters of the gates and fidelity between the gate output state and perfect odd Schrödinger cat state for the fixed half-spacing of 2.87 between the cat state components.}\label{Tab1}
		\centering
		\begin{tabular}{|m{2cm}|m{1.8cm}|m{1.8cm}|m{1.8cm}|}
			\hline
			\multirow{3}{2cm}{$\rho = \sqrt{3}/2;\newline \tau = 1/2 $} & $\gamma = 0.106$ & $\gamma = 0.197$ & $\gamma = 0.296$ \\
			&$y_{m} = 9.05$& $y_{m} = 16.87$ & $y_{m} = 25.39$ \\
			& $F=	0.9976$& $F= 0.9993$ & $F= 0.9995$\\ 
			\hline \hline
			\multirow{3}{2cm}{$\rho = 1/\sqrt{2};\newline \tau = 1/\sqrt{2} $} & $\gamma = 0.111$ & $\gamma = 0.205$ & $\gamma = 0.299$ \\
			&$y_{m} = 3.87$& $y_{m} = 7.17$ & $y_{m} = 10.45$ \\
			& $F=	0.9768$& $F= 0.9933$ & $F= 0.9969$\\ 
			\hline \hline
			\multirow{3}{2cm}{$\rho =1/2;\newline \tau = \sqrt{3}/2$}  
			&$\gamma = 0.115$&$\gamma = 0.202$& $\gamma = 0.289$ \\
			&$y_{m} = 1.89$&$y_{m} = 3.34$& $y_{m} = 4.77$ \\ 
			&$F=	0.7742$ &$F= 0.9244$&$F= 0.9626$\\
			\hline
		\end{tabular}
	\end{table}

        We have demonstrated in Sec.~\ref{sec_semicl} a clear visual interpretation of the gate operation that illustrates the gate fidelity and probability of success in dependence on the cubic nonlinearity and on the measurement outcome. In Fig.~\ref{figCats}, we show a set of the Wigner functions for the same gates, as in Table~\ref{Tab1}, whose properties are in full agreement with our analysis. 
        As far as the half-spacing between the centers of the copies is fixed, the larger the value of the measurement outcome $y_{m}$, the higher than the ``bottom'' of parabola the crossings are located. Hence, the branches of the parabola tend to become parallel in a relevant region, the shearing deformation vanishes in Fig.~\ref{figCats}, and the fidelity increases. 

        In the semiclassical picture, the copies squeezing in coordinate is given (\ref{shifts}) by the squeezing factor $s=1/\rho$. As the beamsplitter reflection coefficient increased, the copies become more squeezed, as illustrated in Fig.~\ref{figCats}. A side effect of this is that the shearing deformation become more apparent. All in all, an optimization of the gate implies a compromise between the tendencies illustrated in Fig.~\ref{figCats}.
		
	\section{Comparison of beamsplitter-based gates: cubic phase resource state versus Fock state}          \label{secComparison}

        To date, several Schr\"odinger cat state generation protocols have been proposed and implemented, which are based on entangling of the input channels and use non-Gaussian states as a resource. 
        For instance, the photon-number states were taken (\cite{Lund2004,Ourjoumtsev2007,Sychev2017})  as a resource having in mind their extensive research and experimental feasibility. In this section, we compare  the beamsplitter-based gates using different resource states, namely, cubic phase state and Fock state.

        Since the squeezed Schr\"odinger cat states \cite{Schlegel2022} are of particular interest for quantum computing and quantum error correction, we examine here the case of an asymmetric beamsplitter with $\rho=1/2$ and $\tau=\sqrt{3}/2$, which  provides squeezing of $s=2$. 
        
        Considering the Fock state-based gate, one can repeat the same semiclassical reasoning that led us to the quadrature amplitudes \eqref{shifts} of the output signal. A vacuum input state is implied,
        and instead of last two lines of (\ref{NonG_amplitudes}) we have for the 
        Fock resource state in semiclassical approximation, 
            \BE
		\big(q_{2}^{(\text{in})}(0)\big)^2 + \big(p_{2}^{(\text{in})}(0)\big)^2 = 2n + 1,
	    \EE
        where $q_{2}^{(\text{in})}(0), p_{2}^{(\text{in})}(0)$ are quadrature amplitudes of the resource state prior to the entangling operation. Here $n$ is the number of photons in the resource state.     
        Then, after the beamsplitter transformation described by the same unitary matrix $S$ \eqref{Smatr} and followed by the momentum measurement of the resource field with the outcome $y_{m}$, we obtain
            \BEA 
            \label{shiftsFock}
		\begin{split}
		&{q}_{1}^{(F)}=\rho {q}_{1}^{(\text{in})}(0) \mp \tau \left[2n+1 - (y_{m} - \tau  {p}_{1}^{(\text{in})}(0))^{2}/\rho^{2} \right]^{1/2},\\
		&{p}_{1}^{(F)}= (1/\rho){p}_{1}^{(\text{in})}(0) - \tau y_{m}/\rho.
		\end{split}
		\EEA
        The output state components are squeezed just as in the cubic phase state-based gate, compare \eqref{shiftsFock} and \eqref{shifts}. The nonlinearity of mapping of the signal amplitudes is evidently different and reflects a different shape of the resource.
       
\begin{figure}[t]
			\begin{center}
				\includegraphics[width=0.9\linewidth]{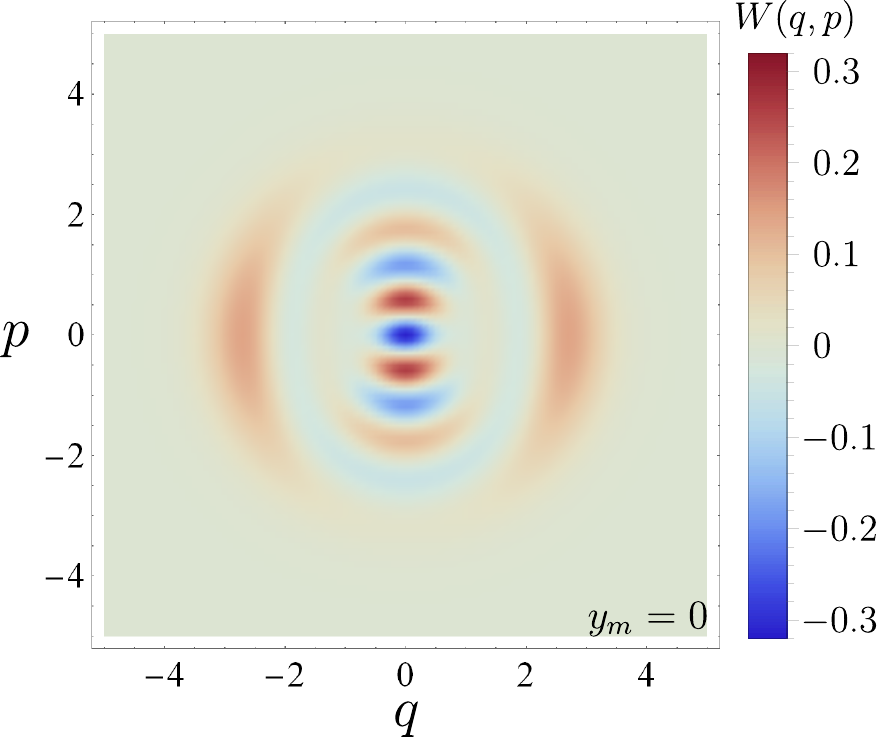}
			\end{center}
			\caption{Wigner function of the output Schrödinger cat state for a Fock state-based gate with an asymmetric beamsplitter $\rho =1/2; \; \tau =\sqrt{3}/2$ and number of photons $n=5$. The optimal measurement outcome is $y_{m}=0$.}
			\label{figCatsFock}
		\end{figure}

        In what follows, we fix an optimal value of $y_m=0$ of the momentum measurement outcome. It is clear from \eqref{shiftsFock} that  characteristic bends in the Wigner function of the output state shown in Fig.~\ref{figCatsFock} are due to the dependence of the target state coordinate $q_1^{(F)}$ on the initial momentum $p_1^{(in)}(0)$.

        In order to evaluate the quality of the Fock state-based gate output numerically, we calculate the fidelity between its  output state and the perfect odd squeezed Schrödinger cat state \eqref{IdCat}. In analogy to Sec.~\ref{GateDef}, the unnormalized wave function of the output state is
            \BEA
            \label{PsiIntOutFock}
				\begin{split}
				\widetilde{\psi}^{(out, n)}_{y_{m}}(x_{1})= \frac{1}{\sqrt{2\pi}}  \int\limits_{-\infty}^{+\infty} dx_{2} \; \psi_{1}(\rho  x_{1} + \tau  x_{2}) \times \\ 
				\times \psi^{(n)}_{2}(-\tau  x_{1} + \rho  x_{2})  e^{-i x_{2} y_{m}},
			\end{split}
		\EEA
        where the wave function of the Fock state with $n$ photons has the form
            \BE
			\psi^{(n)}(x) = \frac{1}{\pi^{1/4}\sqrt{2^n n!}} H_n(x)e^{-x^{2}/2}.\nn
		\EE
        Recall that our perfect odd squeezed Schrödinger cat \eqref{IdCat} is characterized by the relative phase $\theta = \pi$ between the cat components, the half-spacing $\Delta q = \alpha' = \tau \sqrt{2n+1}$  between the copy centers, and the position $\alpha'' = \tau y_{m}/\rho$ in the phase plane. The values of $\alpha',\,\alpha''$ are estimated using \eqref{shiftsFock} for $q_{1}^{(in)}(0)=0,\, p_{1}^{(in)}(0)=0$.

        For the Fock state-based gate with the photon number $n = 5$, one finds $\Delta q = 2.87$ for an optimal measurement outcome  $y_{m}=0$, when the Schrödinger cat state with the squeezing $s'=1/\rho=2$ arises.
        
        Following the same steps as in Sec.~\ref{GateDef}, we evaluate the Wigner function of the gate output, shown in Fig.~\ref{figCatsFock}, and the gate fidelity. To be specific, we consider here a beamsplitter with $\rho=1/2,\,\tau=\sqrt{3}/2$. The fidelity is found to be $F=0.865$, which is close to that for $\gamma=0.115$ in the case of the cubic phase state-base gate.                 
        The nonlinear deformations for both gates are significant for the chosen parameters, thus facilitating the qualitative comparison of the gates.
        
         For the cubic phase state-based scheme, the dominant effect is the shearing deformation of the copies in the phase plane. This is directly related to the slope of branches of the parabola that represents the resource state in Fig.~\ref{figBS}. To improve the fidelity, it is essential to increase the cubic  nonlinearity, which has proven to be a challenging task to date. The maximum value of cubic nonlinearity achieved in the experiment \cite{Eriksson2024} corresponds to $\gamma=0.11$.
         
         The bending deformation produced by the  Fock state-based gate is in a similar way associated with the shape of the resource state in the phase plane. One can reduce the deformation by increasing number of photons in the resource state, and hence improve the fidelity. An essential effect of increasing the photon number on the fidelity has been demonstrated \cite{Baeva2024, Veselkova2025} in more detail for the gate that uses a quantum non-demolition operation instead of a beamsplitter. However, an efficient generation of the Fock states with a large number of photons remains a challenging task as well.
         
         Thus, the choice of an optimal resource state and entangling operation depends on the available experimental capabilities and required fidelity.
               	
		\section{Conclusion}
        
        We have shown that the gate based on the cubic phase state as a non-Gaussian resource and the beamsplitter used to entangle an input signal with the resource allows one to effectively generate squeezed Schr\"odinger cat states, which are quantum superpositions of squeezed ``copies'' of an input state. Such states can be effectively used, for example, for quantum error correction in quantum computing in continuous variables. In general, the gate is able to produce squeezed cat states not only from a vacuum input, but from other input states.

        The semiclassical description of our gate allowed us to predict and visualize possible deformations of the output state as compared with a ``perfect'' squeezed cat state. We have revealed how these imperfections are related to both the shape of the resource state and the choice of beamsplitter reflection and transmission coefficients. 
        
        It is demonstrated that for the cubic phase state-based gate, main imperfections are associated with the shearing deformations of the copies and with the quantum noise which stems from a finite initial squeezing of the resource. To improve the fidelity, one has to increase the degree of cubic nonlinearity in the system and to optimize initial resource squeezing. 

        By comparing in the semiclassical picture the cubic phase state-based gate with its Fock state-based analog, we have found that both devices squeeze the cat state components in the same way. A different shape of the photon number-state manifests itself in a bend of the copies.
        
        In conclusion, one can consider the gate described in our work as one of promising tools for generation of the squeezed Schr\"odinger cat states. Eventually, choosing an optimal resource state requires a balance between experimental availability and theoretical predictions.
			
	\section*{Acknowledgments}
    
        A. V. B. acknowledges a financial support from the Foundation for the Advancement of Theoretical Physics and Mathematics “BASIS” (Grant No. 23-1-5-116-1).
    
        \section*{Author contributions}
        
        A.V.B. performed the calculations with the assistance of A.S.L. All authors contributed to the analysis of results and the preparation of the manuscript. 
        
        \begin{center} 
        Data Availability Statement 
        \end{center}
        
        This manuscript has no associated data or the data will not be deposited.

	\bibliography{Bibliography.bib}

\begin{thebibliography}{33}%
\makeatletter
\providecommand \@ifxundefined [1]{%
 \@ifx{#1\undefined}
}%
\providecommand \@ifnum [1]{%
 \ifnum #1\expandafter \@firstoftwo
 \else \expandafter \@secondoftwo
 \fi
}%
\providecommand \@ifx [1]{%
 \ifx #1\expandafter \@firstoftwo
 \else \expandafter \@secondoftwo
 \fi
}%
\providecommand \natexlab [1]{#1}%
\providecommand \enquote  [1]{``#1''}%
\providecommand \bibnamefont  [1]{#1}%
\providecommand \bibfnamefont [1]{#1}%
\providecommand \citenamefont [1]{#1}%
\providecommand \href@noop [0]{\@secondoftwo}%
\providecommand \href [0]{\begingroup \@sanitize@url \@href}%
\providecommand \@href[1]{\@@startlink{#1}\@@href}%
\providecommand \@@href[1]{\endgroup#1\@@endlink}%
\providecommand \@sanitize@url [0]{\catcode `\\12\catcode `\$12\catcode
  `\&12\catcode `\#12\catcode `\^12\catcode `\_12\catcode `\%12\relax}%
\providecommand \@@startlink[1]{}%
\providecommand \@@endlink[0]{}%
\providecommand \url  [0]{\begingroup\@sanitize@url \@url }%
\providecommand \@url [1]{\endgroup\@href {#1}{\urlprefix }}%
\providecommand \urlprefix  [0]{URL }%
\providecommand \Eprint [0]{\href }%
\providecommand \doibase [0]{https://doi.org/}%
\providecommand \selectlanguage [0]{\@gobble}%
\providecommand \bibinfo  [0]{\@secondoftwo}%
\providecommand \bibfield  [0]{\@secondoftwo}%
\providecommand \translation [1]{[#1]}%
\providecommand \BibitemOpen [0]{}%
\providecommand \bibitemStop [0]{}%
\providecommand \bibitemNoStop [0]{.\EOS\space}%
\providecommand \EOS [0]{\spacefactor3000\relax}%
\providecommand \BibitemShut  [1]{\csname bibitem#1\endcsname}%
\let\auto@bib@innerbib\@empty
\bibitem [{\citenamefont {Yurke}\ and\ \citenamefont
  {Stoler}(1986)}]{Yurke1986}%
  \BibitemOpen
  \bibfield  {author} {\bibinfo {author} {\bibfnamefont {B.}~\bibnamefont
  {Yurke}}\ and\ \bibinfo {author} {\bibfnamefont {D.}~\bibnamefont {Stoler}},\
  }\bibfield  {title} {\bibinfo {title} {Generating quantum mechanical
  superpositions of macroscopically distinguishable states via amplitude
  dispersion},\ }\href {https://doi.org/10.1103/PhysRevLett.57.13} {\bibfield
  {journal} {\bibinfo  {journal} {Phys. Rev. Lett.}\ }\textbf {\bibinfo
  {volume} {57}},\ \bibinfo {pages} {13} (\bibinfo {year} {1986})}\BibitemShut
  {NoStop}%
\bibitem [{\citenamefont {Gerry}\ and\ \citenamefont
  {Knight}(1997)}]{Gerry-Knight-1997}%
  \BibitemOpen
  \bibfield  {author} {\bibinfo {author} {\bibfnamefont {C.~C.}\ \bibnamefont
  {Gerry}}\ and\ \bibinfo {author} {\bibfnamefont {P.~L.}\ \bibnamefont
  {Knight}},\ }\bibfield  {title} {\bibinfo {title} {Quantum superpositions and
  {S}chrödinger cat states in quantum optics},\ }\href
  {https://doi.org/10.1119/1.18698} {\bibfield  {journal} {\bibinfo  {journal}
  {American Journal of Physics}\ }\textbf {\bibinfo {volume} {65}},\ \bibinfo
  {pages} {964} (\bibinfo {year} {1997})}\BibitemShut {NoStop}%
\bibitem [{\citenamefont {Cochrane}\ \emph {et~al.}(1999)\citenamefont
  {Cochrane}, \citenamefont {Milburn},\ and\ \citenamefont
  {Munro}}]{Cochrane1999}%
  \BibitemOpen
  \bibfield  {author} {\bibinfo {author} {\bibfnamefont {P.~T.}\ \bibnamefont
  {Cochrane}}, \bibinfo {author} {\bibfnamefont {G.~J.}\ \bibnamefont
  {Milburn}},\ and\ \bibinfo {author} {\bibfnamefont {W.~J.}\ \bibnamefont
  {Munro}},\ }\bibfield  {title} {\bibinfo {title} {Macroscopically distinct
  quantum-superposition states as a bosonic code for amplitude damping},\
  }\href {https://doi.org/10.1103/PhysRevA.59.2631} {\bibfield  {journal}
  {\bibinfo  {journal} {Phys. Rev. A}\ }\textbf {\bibinfo {volume} {59}},\
  \bibinfo {pages} {2631} (\bibinfo {year} {1999})}\BibitemShut {NoStop}%
\bibitem [{\citenamefont {Leggett}(2002)}]{Leggett2002}%
  \BibitemOpen
  \bibfield  {author} {\bibinfo {author} {\bibfnamefont {A.~J.}\ \bibnamefont
  {Leggett}},\ }\bibfield  {title} {\bibinfo {title} {Testing the limits of
  quantum mechanics: motivation, state of play, prospects},\ }\href
  {https://doi.org/10.1088/0953-8984/14/15/201} {\bibfield  {journal} {\bibinfo
   {journal} {Journal of Physics: Condensed Matter}\ }\textbf {\bibinfo
  {volume} {14}},\ \bibinfo {pages} {R415} (\bibinfo {year}
  {2002})}\BibitemShut {NoStop}%
\bibitem [{\citenamefont {Arndt}\ and\ \citenamefont
  {Hornberger}(2014)}]{Arndt2014}%
  \BibitemOpen
  \bibfield  {author} {\bibinfo {author} {\bibfnamefont {M.}~\bibnamefont
  {Arndt}}\ and\ \bibinfo {author} {\bibfnamefont {K.}~\bibnamefont
  {Hornberger}},\ }\bibfield  {title} {\bibinfo {title} {Testing the limits of
  quantum mechanical superpositions},\ }\href
  {https://doi.org/10.1038/nphys2863} {\bibfield  {journal} {\bibinfo
  {journal} {Nature Physics}\ }\textbf {\bibinfo {volume} {10}},\ \bibinfo
  {pages} {271} (\bibinfo {year} {2014})}\BibitemShut {NoStop}%
\bibitem [{\citenamefont {Gilchrist}\ \emph {et~al.}(2004)\citenamefont
  {Gilchrist}, \citenamefont {Nemoto}, \citenamefont {Munro}, \citenamefont
  {Ralph}, \citenamefont {Glancy}, \citenamefont {Braunstein},\ and\
  \citenamefont {Milburn}}]{Gilchrist2004}%
  \BibitemOpen
  \bibfield  {author} {\bibinfo {author} {\bibfnamefont {A.}~\bibnamefont
  {Gilchrist}}, \bibinfo {author} {\bibfnamefont {K.}~\bibnamefont {Nemoto}},
  \bibinfo {author} {\bibfnamefont {W.~J.}\ \bibnamefont {Munro}}, \bibinfo
  {author} {\bibfnamefont {T.~C.}\ \bibnamefont {Ralph}}, \bibinfo {author}
  {\bibfnamefont {S.}~\bibnamefont {Glancy}}, \bibinfo {author} {\bibfnamefont
  {S.~L.}\ \bibnamefont {Braunstein}},\ and\ \bibinfo {author} {\bibfnamefont
  {G.~J.}\ \bibnamefont {Milburn}},\ }\bibfield  {title} {\bibinfo {title}
  {{S}chr\"odinger cats and their power for quantum information processing},\
  }\href {https://doi.org/10.1088/1464-4266/6/8/032} {\bibfield  {journal}
  {\bibinfo  {journal} {Journal of Optics B: Quantum and Semiclassical Optics}\
  }\textbf {\bibinfo {volume} {6}},\ \bibinfo {pages} {S828} (\bibinfo {year}
  {2004})}\BibitemShut {NoStop}%
\bibitem [{\citenamefont {Giovannetti}\ \emph {et~al.}(2011)\citenamefont
  {Giovannetti}, \citenamefont {Lloyd},\ and\ \citenamefont
  {Maccone}}]{Giovannetti2011}%
  \BibitemOpen
  \bibfield  {author} {\bibinfo {author} {\bibfnamefont {V.}~\bibnamefont
  {Giovannetti}}, \bibinfo {author} {\bibfnamefont {S.}~\bibnamefont {Lloyd}},\
  and\ \bibinfo {author} {\bibfnamefont {L.}~\bibnamefont {Maccone}},\
  }\bibfield  {title} {\bibinfo {title} {Advances in quantum metrology},\
  }\href {https://doi.org/10.1038/nphoton.2011.35} {\bibfield  {journal}
  {\bibinfo  {journal} {Nature Photonics}\ }\textbf {\bibinfo {volume} {5}},\
  \bibinfo {pages} {222} (\bibinfo {year} {2011})}\BibitemShut {NoStop}%
\bibitem [{\citenamefont {Ralph}\ \emph {et~al.}(2003)\citenamefont {Ralph},
  \citenamefont {Gilchrist}, \citenamefont {Milburn}, \citenamefont {Munro},\
  and\ \citenamefont {Glancy}}]{Ralph2003}%
  \BibitemOpen
  \bibfield  {author} {\bibinfo {author} {\bibfnamefont {T.~C.}\ \bibnamefont
  {Ralph}}, \bibinfo {author} {\bibfnamefont {A.}~\bibnamefont {Gilchrist}},
  \bibinfo {author} {\bibfnamefont {G.~J.}\ \bibnamefont {Milburn}}, \bibinfo
  {author} {\bibfnamefont {W.~J.}\ \bibnamefont {Munro}},\ and\ \bibinfo
  {author} {\bibfnamefont {S.}~\bibnamefont {Glancy}},\ }\bibfield  {title}
  {\bibinfo {title} {Quantum computation with optical coherent states},\ }\href
  {https://doi.org/10.1103/PhysRevA.68.042319} {\bibfield  {journal} {\bibinfo
  {journal} {Phys. Rev. A}\ }\textbf {\bibinfo {volume} {68}},\ \bibinfo
  {pages} {042319} (\bibinfo {year} {2003})}\BibitemShut {NoStop}%
\bibitem [{\citenamefont {Lund}\ \emph {et~al.}(2008)\citenamefont {Lund},
  \citenamefont {Ralph},\ and\ \citenamefont {Haselgrove}}]{Lund2008}%
  \BibitemOpen
  \bibfield  {author} {\bibinfo {author} {\bibfnamefont {A.~P.}\ \bibnamefont
  {Lund}}, \bibinfo {author} {\bibfnamefont {T.~C.}\ \bibnamefont {Ralph}},\
  and\ \bibinfo {author} {\bibfnamefont {H.~L.}\ \bibnamefont {Haselgrove}},\
  }\bibfield  {title} {\bibinfo {title} {Fault-tolerant linear optical quantum
  computing with small-amplitude coherent states},\ }\href
  {https://doi.org/10.1103/PhysRevLett.100.030503} {\bibfield  {journal}
  {\bibinfo  {journal} {Phys. Rev. Lett.}\ }\textbf {\bibinfo {volume} {100}},\
  \bibinfo {pages} {030503} (\bibinfo {year} {2008})}\BibitemShut {NoStop}%
\bibitem [{\citenamefont {Guillaud}\ \emph {et~al.}(2023)\citenamefont
  {Guillaud}, \citenamefont {Cohen},\ and\ \citenamefont
  {Mirrahimi}}]{Guillaud2023}%
  \BibitemOpen
  \bibfield  {author} {\bibinfo {author} {\bibfnamefont {J.}~\bibnamefont
  {Guillaud}}, \bibinfo {author} {\bibfnamefont {J.}~\bibnamefont {Cohen}},\
  and\ \bibinfo {author} {\bibfnamefont {M.}~\bibnamefont {Mirrahimi}},\
  }\bibfield  {title} {\bibinfo {title} {Quantum computation with cat qubits},\
  }\href {https://doi.org/10.21468/SciPostPhysLectNotes.72} {\bibfield
  {journal} {\bibinfo  {journal} {SciPost Physics Lecture Notes}\ ,\ \bibinfo
  {pages} {1}} (\bibinfo {year} {2023})}\BibitemShut {NoStop}%
\bibitem [{\citenamefont {Le~Jeannic}\ \emph {et~al.}(2018)\citenamefont
  {Le~Jeannic}, \citenamefont {Cavaillès}, \citenamefont {Huang},
  \citenamefont {Filip},\ and\ \citenamefont {Laurat}}]{LeJeannic2018}%
  \BibitemOpen
  \bibfield  {author} {\bibinfo {author} {\bibfnamefont {H.}~\bibnamefont
  {Le~Jeannic}}, \bibinfo {author} {\bibfnamefont {A.}~\bibnamefont
  {Cavaillès}}, \bibinfo {author} {\bibfnamefont {K.}~\bibnamefont {Huang}},
  \bibinfo {author} {\bibfnamefont {R.}~\bibnamefont {Filip}},\ and\ \bibinfo
  {author} {\bibfnamefont {J.}~\bibnamefont {Laurat}},\ }\bibfield  {title}
  {\bibinfo {title} {Slowing quantum decoherence by squeezing in phase space},\
  }\bibfield  {journal} {\bibinfo  {journal} {Physical Review Letters}\
  }\textbf {\bibinfo {volume} {120}},\ \href
  {https://doi.org/10.1103/physrevlett.120.073603}
  {10.1103/physrevlett.120.073603} (\bibinfo {year} {2018})\BibitemShut
  {NoStop}%
\bibitem [{\citenamefont {Schlegel}\ \emph {et~al.}(2022)\citenamefont
  {Schlegel}, \citenamefont {Minganti},\ and\ \citenamefont
  {Savona}}]{Schlegel2022}%
  \BibitemOpen
  \bibfield  {author} {\bibinfo {author} {\bibfnamefont {D.~S.}\ \bibnamefont
  {Schlegel}}, \bibinfo {author} {\bibfnamefont {F.}~\bibnamefont {Minganti}},\
  and\ \bibinfo {author} {\bibfnamefont {V.}~\bibnamefont {Savona}},\
  }\bibfield  {title} {\bibinfo {title} {Quantum error correction using
  squeezed {S}chr\"odinger cat states},\ }\href
  {https://doi.org/10.1103/PhysRevA.106.022431} {\bibfield  {journal} {\bibinfo
   {journal} {Phys. Rev. A}\ }\textbf {\bibinfo {volume} {106}},\ \bibinfo
  {pages} {022431} (\bibinfo {year} {2022})}\BibitemShut {NoStop}%
\bibitem [{\citenamefont {Rivera-Dean}\ \emph {et~al.}(2021)\citenamefont
  {Rivera-Dean}, \citenamefont {Stammer}, \citenamefont {Pisanty},
  \citenamefont {Lamprou}, \citenamefont {Tzallas}, \citenamefont
  {Lewenstein},\ and\ \citenamefont {Ciappina}}]{RiveraDean2021}%
  \BibitemOpen
  \bibfield  {author} {\bibinfo {author} {\bibfnamefont {J.}~\bibnamefont
  {Rivera-Dean}}, \bibinfo {author} {\bibfnamefont {P.}~\bibnamefont
  {Stammer}}, \bibinfo {author} {\bibfnamefont {E.}~\bibnamefont {Pisanty}},
  \bibinfo {author} {\bibfnamefont {T.}~\bibnamefont {Lamprou}}, \bibinfo
  {author} {\bibfnamefont {P.}~\bibnamefont {Tzallas}}, \bibinfo {author}
  {\bibfnamefont {M.}~\bibnamefont {Lewenstein}},\ and\ \bibinfo {author}
  {\bibfnamefont {M.~F.}\ \bibnamefont {Ciappina}},\ }\bibfield  {title}
  {\bibinfo {title} {New schemes for creating large optical {S}chr\"{o}dinger
  cat states using strong laser fields},\ }\href
  {https://doi.org/10.1007/s10825-021-01789-2} {\bibfield  {journal} {\bibinfo
  {journal} {Journal of Computational Electronics}\ }\textbf {\bibinfo {volume}
  {20}},\ \bibinfo {pages} {2111} (\bibinfo {year} {2021})}\BibitemShut
  {NoStop}%
\bibitem [{\citenamefont {Ourjoumtsev}\ \emph {et~al.}(2007)\citenamefont
  {Ourjoumtsev}, \citenamefont {Tualle-Brouri},\ and\ \citenamefont
  {Grangier}}]{Ourjoumtsev2007}%
  \BibitemOpen
  \bibfield  {author} {\bibinfo {author} {\bibfnamefont {A.}~\bibnamefont
  {Ourjoumtsev}}, \bibinfo {author} {\bibfnamefont {R.}~\bibnamefont
  {Tualle-Brouri}},\ and\ \bibinfo {author} {\bibfnamefont {P.}~\bibnamefont
  {Grangier}},\ }\bibfield  {title} {\bibinfo {title} {Generation of optical
  <<{S}chr\"odinger cats>> from photon number states},\ }\href
  {https://doi.org/10.1038/nature06054} {\bibfield  {journal} {\bibinfo
  {journal} {Nature}\ }\textbf {\bibinfo {volume} {448}},\ \bibinfo {pages}
  {784} (\bibinfo {year} {2007})}\BibitemShut {NoStop}%
\bibitem [{\citenamefont {Yurke}\ \emph {et~al.}(1990)\citenamefont {Yurke},
  \citenamefont {Schleich},\ and\ \citenamefont {Walls}}]{Yurke1990}%
  \BibitemOpen
  \bibfield  {author} {\bibinfo {author} {\bibfnamefont {B.}~\bibnamefont
  {Yurke}}, \bibinfo {author} {\bibfnamefont {W.}~\bibnamefont {Schleich}},\
  and\ \bibinfo {author} {\bibfnamefont {D.~F.}\ \bibnamefont {Walls}},\
  }\bibfield  {title} {\bibinfo {title} {Quantum superpositions generated by
  quantum nondemolition measurements},\ }\href
  {https://doi.org/10.1103/PhysRevA.42.1703} {\bibfield  {journal} {\bibinfo
  {journal} {Phys. Rev. A}\ }\textbf {\bibinfo {volume} {42}},\ \bibinfo
  {pages} {1703} (\bibinfo {year} {1990})}\BibitemShut {NoStop}%
\bibitem [{\citenamefont {Korolev}\ \emph {et~al.}(2024)\citenamefont
  {Korolev}, \citenamefont {Bashmakova},\ and\ \citenamefont
  {Golubeva}}]{Korolev2024}%
  \BibitemOpen
  \bibfield  {author} {\bibinfo {author} {\bibfnamefont {S.~B.}\ \bibnamefont
  {Korolev}}, \bibinfo {author} {\bibfnamefont {E.~N.}\ \bibnamefont
  {Bashmakova}},\ and\ \bibinfo {author} {\bibfnamefont {T.~Y.}\ \bibnamefont
  {Golubeva}},\ }\bibfield  {title} {\bibinfo {title} {Error correction using
  squeezed fock states},\ }\bibfield  {journal} {\bibinfo  {journal} {Quantum
  Information Processing}\ }\textbf {\bibinfo {volume} {23}},\ \href
  {https://doi.org/10.1007/s11128-024-04549-w} {10.1007/s11128-024-04549-w}
  (\bibinfo {year} {2024})\BibitemShut {NoStop}%
\bibitem [{\citenamefont {Ourjoumtsev}\ \emph {et~al.}(2006)\citenamefont
  {Ourjoumtsev}, \citenamefont {Tualle-Brouri}, \citenamefont {Laurat},\ and\
  \citenamefont {Grangier}}]{Ourjoumtsev2006}%
  \BibitemOpen
  \bibfield  {author} {\bibinfo {author} {\bibfnamefont {A.}~\bibnamefont
  {Ourjoumtsev}}, \bibinfo {author} {\bibfnamefont {R.}~\bibnamefont
  {Tualle-Brouri}}, \bibinfo {author} {\bibfnamefont {J.}~\bibnamefont
  {Laurat}},\ and\ \bibinfo {author} {\bibfnamefont {P.}~\bibnamefont
  {Grangier}},\ }\bibfield  {title} {\bibinfo {title} {Generating optical
  {S}chr\"odinger kittens for quantum information processing},\ }\href
  {https://doi.org/10.1126/science.1122858} {\bibfield  {journal} {\bibinfo
  {journal} {Science}\ }\textbf {\bibinfo {volume} {312}},\ \bibinfo {pages}
  {83} (\bibinfo {year} {2006})}\BibitemShut {NoStop}%
\bibitem [{\citenamefont {Takase}\ \emph {et~al.}(2022)\citenamefont {Takase},
  \citenamefont {Kawasaki}, \citenamefont {Jeong}, \citenamefont {Endo},
  \citenamefont {Kashiwazaki}, \citenamefont {Kazama}, \citenamefont {Enbutsu},
  \citenamefont {Watanabe}, \citenamefont {Umeki}, \citenamefont {Miki},
  \citenamefont {Terai}, \citenamefont {Yabuno}, \citenamefont {China},
  \citenamefont {Asavanant}, \citenamefont {Yoshikawa},\ and\ \citenamefont
  {Furusawa}}]{Takase2022}%
  \BibitemOpen
  \bibfield  {author} {\bibinfo {author} {\bibfnamefont {K.}~\bibnamefont
  {Takase}}, \bibinfo {author} {\bibfnamefont {A.}~\bibnamefont {Kawasaki}},
  \bibinfo {author} {\bibfnamefont {B.~K.}\ \bibnamefont {Jeong}}, \bibinfo
  {author} {\bibfnamefont {M.}~\bibnamefont {Endo}}, \bibinfo {author}
  {\bibfnamefont {T.}~\bibnamefont {Kashiwazaki}}, \bibinfo {author}
  {\bibfnamefont {T.}~\bibnamefont {Kazama}}, \bibinfo {author} {\bibfnamefont
  {K.}~\bibnamefont {Enbutsu}}, \bibinfo {author} {\bibfnamefont
  {K.}~\bibnamefont {Watanabe}}, \bibinfo {author} {\bibfnamefont
  {T.}~\bibnamefont {Umeki}}, \bibinfo {author} {\bibfnamefont
  {S.}~\bibnamefont {Miki}}, \bibinfo {author} {\bibfnamefont {H.}~\bibnamefont
  {Terai}}, \bibinfo {author} {\bibfnamefont {M.}~\bibnamefont {Yabuno}},
  \bibinfo {author} {\bibfnamefont {F.}~\bibnamefont {China}}, \bibinfo
  {author} {\bibfnamefont {W.}~\bibnamefont {Asavanant}}, \bibinfo {author}
  {\bibfnamefont {J.-i.}\ \bibnamefont {Yoshikawa}},\ and\ \bibinfo {author}
  {\bibfnamefont {A.}~\bibnamefont {Furusawa}},\ }\bibfield  {title} {\bibinfo
  {title} {Generation of {S}chr\"{o}dinger cat states with {W}igner negativity
  using a continuous-wave low-loss waveguide optical parametric amplifier},\
  }\href {https://doi.org/10.1364/oe.454123} {\bibfield  {journal} {\bibinfo
  {journal} {Optics Express}\ }\textbf {\bibinfo {volume} {30}},\ \bibinfo
  {pages} {14161} (\bibinfo {year} {2022})}\BibitemShut {NoStop}%
\bibitem [{\citenamefont {Thekkadath}\ \emph {et~al.}(2020)\citenamefont
  {Thekkadath}, \citenamefont {Bell}, \citenamefont {Walmsley},\ and\
  \citenamefont {Lvovsky}}]{Thekkadath2020}%
  \BibitemOpen
  \bibfield  {author} {\bibinfo {author} {\bibfnamefont {G.~S.}\ \bibnamefont
  {Thekkadath}}, \bibinfo {author} {\bibfnamefont {B.~A.}\ \bibnamefont
  {Bell}}, \bibinfo {author} {\bibfnamefont {I.~A.}\ \bibnamefont {Walmsley}},\
  and\ \bibinfo {author} {\bibfnamefont {A.~I.}\ \bibnamefont {Lvovsky}},\
  }\bibfield  {title} {\bibinfo {title} {Engineering {S}chrödinger cat states
  with a photonic even-parity detector},\ }\href
  {https://doi.org/10.22331/q-2020-03-02-239} {\bibfield  {journal} {\bibinfo
  {journal} {Quantum}\ }\textbf {\bibinfo {volume} {4}},\ \bibinfo {pages}
  {239} (\bibinfo {year} {2020})}\BibitemShut {NoStop}%
\bibitem [{\citenamefont {Lund}\ \emph {et~al.}(2004)\citenamefont {Lund},
  \citenamefont {Jeong}, \citenamefont {Ralph},\ and\ \citenamefont
  {Kim}}]{Lund2004}%
  \BibitemOpen
  \bibfield  {author} {\bibinfo {author} {\bibfnamefont {A.~P.}\ \bibnamefont
  {Lund}}, \bibinfo {author} {\bibfnamefont {H.}~\bibnamefont {Jeong}},
  \bibinfo {author} {\bibfnamefont {T.~C.}\ \bibnamefont {Ralph}},\ and\
  \bibinfo {author} {\bibfnamefont {M.~S.}\ \bibnamefont {Kim}},\ }\bibfield
  {title} {\bibinfo {title} {Conditional production of superpositions of
  coherent states with inefficient photon detection},\ }\href
  {https://doi.org/10.1103/PhysRevA.70.020101} {\bibfield  {journal} {\bibinfo
  {journal} {Phys. Rev. A}\ }\textbf {\bibinfo {volume} {70}},\ \bibinfo
  {pages} {020101(R)} (\bibinfo {year} {2004})}\BibitemShut {NoStop}%
\bibitem [{\citenamefont {Sychev}\ \emph {et~al.}(2017)\citenamefont {Sychev},
  \citenamefont {Ulanov}, \citenamefont {Pushkina}, \citenamefont {Richards},
  \citenamefont {Fedorov},\ and\ \citenamefont {Lvovsky}}]{Sychev2017}%
  \BibitemOpen
  \bibfield  {author} {\bibinfo {author} {\bibfnamefont {D.~V.}\ \bibnamefont
  {Sychev}}, \bibinfo {author} {\bibfnamefont {A.~E.}\ \bibnamefont {Ulanov}},
  \bibinfo {author} {\bibfnamefont {A.~A.}\ \bibnamefont {Pushkina}}, \bibinfo
  {author} {\bibfnamefont {M.~W.}\ \bibnamefont {Richards}}, \bibinfo {author}
  {\bibfnamefont {I.~A.}\ \bibnamefont {Fedorov}},\ and\ \bibinfo {author}
  {\bibfnamefont {A.~I.}\ \bibnamefont {Lvovsky}},\ }\bibfield  {title}
  {\bibinfo {title} {Enlargement of optical {S}chrödinger's cat states},\
  }\href {https://doi.org/10.1038/nphoton.2017.57} {\bibfield  {journal}
  {\bibinfo  {journal} {Nature Photonic}\ }\textbf {\bibinfo {volume} {11}},\
  \bibinfo {pages} {379} (\bibinfo {year} {2017})}\BibitemShut {NoStop}%
\bibitem [{\citenamefont {Eaton}\ \emph {et~al.}(2022)\citenamefont {Eaton},
  \citenamefont {González-Arciniegas}, \citenamefont {Alexander},
  \citenamefont {Menicucci},\ and\ \citenamefont {Pfister}}]{Eaton2022}%
  \BibitemOpen
  \bibfield  {author} {\bibinfo {author} {\bibfnamefont {M.}~\bibnamefont
  {Eaton}}, \bibinfo {author} {\bibfnamefont {C.}~\bibnamefont
  {González-Arciniegas}}, \bibinfo {author} {\bibfnamefont {R.~N.}\
  \bibnamefont {Alexander}}, \bibinfo {author} {\bibfnamefont {N.~C.}\
  \bibnamefont {Menicucci}},\ and\ \bibinfo {author} {\bibfnamefont
  {O.}~\bibnamefont {Pfister}},\ }\bibfield  {title} {\bibinfo {title}
  {Measurement-based generation and preservation of cat and grid states within
  a continuous-variable cluster state},\ }\href
  {https://doi.org/10.22331/q-2022-07-20-769} {\bibfield  {journal} {\bibinfo
  {journal} {Quantum}\ }\textbf {\bibinfo {volume} {6}},\ \bibinfo {pages}
  {769} (\bibinfo {year} {2022})}\BibitemShut {NoStop}%
\bibitem [{\citenamefont {Winnel}\ \emph {et~al.}(2024)\citenamefont {Winnel},
  \citenamefont {Guanzon}, \citenamefont {Singh},\ and\ \citenamefont
  {Ralph}}]{Winnel2024}%
  \BibitemOpen
  \bibfield  {author} {\bibinfo {author} {\bibfnamefont {M.~S.}\ \bibnamefont
  {Winnel}}, \bibinfo {author} {\bibfnamefont {J.~J.}\ \bibnamefont {Guanzon}},
  \bibinfo {author} {\bibfnamefont {D.}~\bibnamefont {Singh}},\ and\ \bibinfo
  {author} {\bibfnamefont {T.~C.}\ \bibnamefont {Ralph}},\ }\bibfield  {title}
  {\bibinfo {title} {Deterministic preparation of optical squeezed cat and
  {G}ottesman-{K}itaev-{P}reskill states},\ }\bibfield  {journal} {\bibinfo
  {journal} {Phys. Rev. Lett.}\ }\textbf {\bibinfo {volume} {132}},\ \href
  {https://doi.org/10.1103/physrevlett.132.230602}
  {10.1103/physrevlett.132.230602} (\bibinfo {year} {2024})\BibitemShut
  {NoStop}%
\bibitem [{\citenamefont {Sokolov}(2020)}]{Sokolov2020}%
  \BibitemOpen
  \bibfield  {author} {\bibinfo {author} {\bibfnamefont {I.}~\bibnamefont
  {Sokolov}},\ }\bibfield  {title} {\bibinfo {title} {Schr\"{o}dinger cat
  states in continuous variable non-{G}aussian networks},\ }\href
  {https://doi.org/10.1016/j.physleta.2020.126762} {\bibfield  {journal}
  {\bibinfo  {journal} {Physics Letters A}\ }\textbf {\bibinfo {volume}
  {384}},\ \bibinfo {pages} {126762} (\bibinfo {year} {2020})}\BibitemShut
  {NoStop}%
\bibitem [{\citenamefont {Masalaeva}\ and\ \citenamefont
  {Sokolov}(2021)}]{Masalaeva2022}%
  \BibitemOpen
  \bibfield  {author} {\bibinfo {author} {\bibfnamefont {N.}~\bibnamefont
  {Masalaeva}}\ and\ \bibinfo {author} {\bibfnamefont {I.}~\bibnamefont
  {Sokolov}},\ }\bibfield  {title} {\bibinfo {title} {Quantum statistics of
  {S}chr\"{o}dinger cat states prepared by logical gate with non-{G}aussian
  resource state},\ }\href {https://doi.org/10.1016/j.physleta.2021.127846}
  {\bibfield  {journal} {\bibinfo  {journal} {Physics Letters A}\ }\textbf
  {\bibinfo {volume} {424}},\ \bibinfo {pages} {127846} (\bibinfo {year}
  {2021})}\BibitemShut {NoStop}%
\bibitem [{\citenamefont {Baeva}\ \emph {et~al.}(2023)\citenamefont {Baeva},
  \citenamefont {Losev},\ and\ \citenamefont {Sokolov}}]{Baeva2023}%
  \BibitemOpen
  \bibfield  {author} {\bibinfo {author} {\bibfnamefont {A.}~\bibnamefont
  {Baeva}}, \bibinfo {author} {\bibfnamefont {A.}~\bibnamefont {Losev}},\ and\
  \bibinfo {author} {\bibfnamefont {I.}~\bibnamefont {Sokolov}},\ }\bibfield
  {title} {\bibinfo {title} {Schr\"{o}dinger cat states prepared by logical
  gate with non-{G}aussian resource state: {E}ffect of finite squeezing and
  efficiency versus monotones},\ }\href
  {https://doi.org/10.1016/j.physleta.2023.128730} {\bibfield  {journal}
  {\bibinfo  {journal} {Physics Letters A}\ }\textbf {\bibinfo {volume}
  {466}},\ \bibinfo {pages} {128730} (\bibinfo {year} {2023})}\BibitemShut
  {NoStop}%
\bibitem [{\citenamefont {Baeva}\ \emph {et~al.}(2024)\citenamefont {Baeva},
  \citenamefont {Veselkova}, \citenamefont {Masalaeva},\ and\ \citenamefont
  {Sokolov}}]{Baeva2024}%
  \BibitemOpen
  \bibfield  {author} {\bibinfo {author} {\bibfnamefont {A.~V.}\ \bibnamefont
  {Baeva}}, \bibinfo {author} {\bibfnamefont {N.~G.}\ \bibnamefont
  {Veselkova}}, \bibinfo {author} {\bibfnamefont {N.~I.}\ \bibnamefont
  {Masalaeva}},\ and\ \bibinfo {author} {\bibfnamefont {I.~V.}\ \bibnamefont
  {Sokolov}},\ }\bibfield  {title} {\bibinfo {title} {Measurement-assisted
  non-{G}aussian gate for {S}chr\"{o}dinger cat states preparation: {F}ock
  resource state versus cubic phase state},\ }\bibfield  {journal} {\bibinfo
  {journal} {The European Physical Journal D}\ }\textbf {\bibinfo {volume}
  {78}},\ \href {https://doi.org/10.1140/epjd/s10053-023-00796-1}
  {10.1140/epjd/s10053-023-00796-1} (\bibinfo {year} {2024})\BibitemShut
  {NoStop}%
\bibitem [{\citenamefont {Braunstein}\ and\ \citenamefont
  {Caves}(1990)}]{Braunstein1990}%
  \BibitemOpen
  \bibfield  {author} {\bibinfo {author} {\bibfnamefont {S.~L.}\ \bibnamefont
  {Braunstein}}\ and\ \bibinfo {author} {\bibfnamefont {C.~M.}\ \bibnamefont
  {Caves}},\ }\bibfield  {title} {\bibinfo {title} {Phase and homodyne
  statistics of generalized squeezed states},\ }\href
  {https://doi.org/10.1103/PhysRevA.42.4115} {\bibfield  {journal} {\bibinfo
  {journal} {Phys. Rev. A}\ }\textbf {\bibinfo {volume} {42}},\ \bibinfo
  {pages} {4115} (\bibinfo {year} {1990})}\BibitemShut {NoStop}%
\bibitem [{\citenamefont {Zelaya}\ \emph {et~al.}(2018)\citenamefont {Zelaya},
  \citenamefont {Dey},\ and\ \citenamefont {Hussin}}]{Zelaya2018}%
  \BibitemOpen
  \bibfield  {author} {\bibinfo {author} {\bibfnamefont {K.}~\bibnamefont
  {Zelaya}}, \bibinfo {author} {\bibfnamefont {S.}~\bibnamefont {Dey}},\ and\
  \bibinfo {author} {\bibfnamefont {V.}~\bibnamefont {Hussin}},\ }\bibfield
  {title} {\bibinfo {title} {Generalized squeezed states},\ }\href
  {https://doi.org/10.1016/j.physleta.2018.10.003} {\bibfield  {journal}
  {\bibinfo  {journal} {Physics Letters A}\ }\textbf {\bibinfo {volume}
  {382}},\ \bibinfo {pages} {3369} (\bibinfo {year} {2018})}\BibitemShut
  {NoStop}%
\bibitem [{\citenamefont {Kudra}\ \emph {et~al.}(2022)\citenamefont {Kudra},
  \citenamefont {Kervinen}, \citenamefont {Strandberg}, \citenamefont {Ahmed},
  \citenamefont {Scigliuzzo}, \citenamefont {Osman}, \citenamefont {Lozano},
  \citenamefont {Thol\'en}, \citenamefont {Borgani}, \citenamefont {Haviland},
  \citenamefont {Ferrini}, \citenamefont {Bylander}, \citenamefont {Kockum},
  \citenamefont {Quijandr\'{\i}a}, \citenamefont {Delsing},\ and\ \citenamefont
  {Gasparinetti}}]{Kudra2022}%
  \BibitemOpen
  \bibfield  {author} {\bibinfo {author} {\bibfnamefont {M.}~\bibnamefont
  {Kudra}}, \bibinfo {author} {\bibfnamefont {M.}~\bibnamefont {Kervinen}},
  \bibinfo {author} {\bibfnamefont {I.}~\bibnamefont {Strandberg}}, \bibinfo
  {author} {\bibfnamefont {S.}~\bibnamefont {Ahmed}}, \bibinfo {author}
  {\bibfnamefont {M.}~\bibnamefont {Scigliuzzo}}, \bibinfo {author}
  {\bibfnamefont {A.}~\bibnamefont {Osman}}, \bibinfo {author} {\bibfnamefont
  {D.~P.}\ \bibnamefont {Lozano}}, \bibinfo {author} {\bibfnamefont {M.~O.}\
  \bibnamefont {Thol\'en}}, \bibinfo {author} {\bibfnamefont {R.}~\bibnamefont
  {Borgani}}, \bibinfo {author} {\bibfnamefont {D.~B.}\ \bibnamefont
  {Haviland}}, \bibinfo {author} {\bibfnamefont {G.}~\bibnamefont {Ferrini}},
  \bibinfo {author} {\bibfnamefont {J.}~\bibnamefont {Bylander}}, \bibinfo
  {author} {\bibfnamefont {A.~F.}\ \bibnamefont {Kockum}}, \bibinfo {author}
  {\bibfnamefont {F.}~\bibnamefont {Quijandr\'{\i}a}}, \bibinfo {author}
  {\bibfnamefont {P.}~\bibnamefont {Delsing}},\ and\ \bibinfo {author}
  {\bibfnamefont {S.}~\bibnamefont {Gasparinetti}},\ }\bibfield  {title}
  {\bibinfo {title} {Robust preparation of {W}igner-negative states with
  optimized {SNAP}-displacement sequences},\ }\href
  {https://doi.org/10.1103/PRXQuantum.3.030301} {\bibfield  {journal} {\bibinfo
   {journal} {PRX Quantum}\ }\textbf {\bibinfo {volume} {3}},\ \bibinfo {pages}
  {030301} (\bibinfo {year} {2022})}\BibitemShut {NoStop}%
\bibitem [{\citenamefont {Eriksson}\ \emph {et~al.}(2024)\citenamefont
  {Eriksson}, \citenamefont {Sépulcre}, \citenamefont {Kervinen},
  \citenamefont {Hillmann}, \citenamefont {Kudra}, \citenamefont {Dupouy},
  \citenamefont {Lu}, \citenamefont {Khanahmadi}, \citenamefont {Yang},
  \citenamefont {Castillo-Moreno}, \citenamefont {Delsing},\ and\ \citenamefont
  {Gasparinetti}}]{Eriksson2024}%
  \BibitemOpen
  \bibfield  {author} {\bibinfo {author} {\bibfnamefont {A.~M.}\ \bibnamefont
  {Eriksson}}, \bibinfo {author} {\bibfnamefont {T.}~\bibnamefont {Sépulcre}},
  \bibinfo {author} {\bibfnamefont {M.}~\bibnamefont {Kervinen}}, \bibinfo
  {author} {\bibfnamefont {T.}~\bibnamefont {Hillmann}}, \bibinfo {author}
  {\bibfnamefont {M.}~\bibnamefont {Kudra}}, \bibinfo {author} {\bibfnamefont
  {S.}~\bibnamefont {Dupouy}}, \bibinfo {author} {\bibfnamefont
  {Y.}~\bibnamefont {Lu}}, \bibinfo {author} {\bibfnamefont {M.}~\bibnamefont
  {Khanahmadi}}, \bibinfo {author} {\bibfnamefont {J.}~\bibnamefont {Yang}},
  \bibinfo {author} {\bibfnamefont {C.}~\bibnamefont {Castillo-Moreno}},
  \bibinfo {author} {\bibfnamefont {P.}~\bibnamefont {Delsing}},\ and\ \bibinfo
  {author} {\bibfnamefont {S.}~\bibnamefont {Gasparinetti}},\ }\bibfield
  {title} {\bibinfo {title} {Universal control of a bosonic mode via
  drive-activated native cubic interactions},\ }\bibfield  {journal} {\bibinfo
  {journal} {Nature Communications}\ }\textbf {\bibinfo {volume} {15}},\ \href
  {https://doi.org/10.1038/s41467-024-46507-1} {10.1038/s41467-024-46507-1}
  (\bibinfo {year} {2024})\BibitemShut {NoStop}%
\bibitem [{\citenamefont {Vall{\'e}e}\ and\ \citenamefont
  {Soares}(2010)}]{vallee2010airy}%
  \BibitemOpen
  \bibfield  {author} {\bibinfo {author} {\bibfnamefont {O.}~\bibnamefont
  {Vall{\'e}e}}\ and\ \bibinfo {author} {\bibfnamefont {M.}~\bibnamefont
  {Soares}},\ }\href@noop {} {\bibinfo {title} {Airy functions and applications
  to physics. 2nd edn. hackensack}} (\bibinfo {year} {2010})\BibitemShut
  {NoStop}%
\bibitem [{\citenamefont {Veselkova}\ \emph {et~al.}(2025)\citenamefont
  {Veselkova}, \citenamefont {Goncharov},\ and\ \citenamefont
  {Kiselev}}]{Veselkova2025}%
  \BibitemOpen
  \bibfield  {author} {\bibinfo {author} {\bibfnamefont {N.~G.}\ \bibnamefont
  {Veselkova}}, \bibinfo {author} {\bibfnamefont {R.}~\bibnamefont
  {Goncharov}},\ and\ \bibinfo {author} {\bibfnamefont {A.~D.}\ \bibnamefont
  {Kiselev}},\ }\href {https://doi.org/10.48550/ARXIV.2501.09318} {\bibinfo
  {title} {Creation and manipulation of schr\"{o}dinger cat states based on
  semiclassical predictions}} (\bibinfo {year} {2025})\BibitemShut {NoStop}%
\end{thebibliography}%
	
\end{document}